\documentstyle[12pt,aaspp4]{article}
\begin{document}
\newcommand{\leteq}{\begin{mathletters}}
\newcommand{\beq}{\end{mathletters}}
\newcommand{\simlt}{\lesssim}
\newcommand{\simgt}{\gtrsim}
\newcommand{\bc}{\begin{center}}
\newcommand{\ec}{\end{center}}
\newcommand{\cc}{\rm{cm}^{-3}}
\newcommand{\ccb}{\rm{cm}^{3}}
\newcommand{\bm}{\boldmath}
\newcommand{\Bvec}{\bm B}
\newcommand{\msol}{M_\odot}
\newcommand{\HII}{\rm{H_2}}
\newcommand{\mHII}{m_{{}_{\HII}}}
\newcommand{\HCO}{\rm{HCO}}
\newcommand{\HCOp}{\rm{HCO^+}}
\newcommand{\Mg}{\rm{Mg}}
\newcommand{\Mgp}{\rm{Mg^+}}
\newcommand{\Na}{\rm{Na}}
\newcommand{\Nap}{\rm{Na^+}}
\newcommand{\Fe}{\rm{Fe}}
\newcommand{\Fep}{\rm{Fe^+}}
\newcommand{\C}{\rm{C}}
\newcommand{\Cp}{\rm{C^+}}
\newcommand{\Su}{\rm{S}}
\newcommand{\Sup}{\rm{S^+}}
\newcommand{\Si}{\rm{Si}}
\newcommand{\Sip}{\rm{Si^+}}
\newcommand{\ul}{\underline{\hspace{40pt}}}
\newcommand{\nn}{n_{\rm n}}
\newcommand{\phiB}{\Phi_B}
\newcommand{\phiBsol}{\Phi_{B,\odot}}
\newcommand{\rsol}{R_{\odot}}
\newcommand{\phiBc}{\Phi_{B,\rm{c}}}
\newcommand{\mi}{m_{\rm i}}
\newcommand{\mn}{m_{\rm n}}
\newcommand{\wi}{\omega_{\rm i}}
\newcommand{\nnc}{n_{\rm{n,c}}}
\newcommand{\nncb}{n_{\rm{n,c0}}}
\newcommand{\sigcb}{\sigma_{\rm{n,c0}}}
\newcommand{\Beq}{B_{z,\rm{eq}}}
\newcommand{\Beqd}{B_{z,\rm{eq,d}}}
\newcommand{\Bequ}{B_{z,\rm{eq,u}}}
\newcommand{\Bref}{B_{\rm{ref}}}
\newcommand{\sigcref}{\sigma_{\rm{c,ref}}}
\newcommand{\BrZ}{B_{r,Z}}
\newcommand{\nni}{n_{\rm i}}
\newcommand{\xxi}{x_{\rm i}}
\newcommand{\nne}{n_{\rm e}}
\newcommand{\nec}{n_{\rm{e,c}}}
\newcommand{\xe}{x_{\rm{e}}}
\newcommand{\phitomd}{\mu_{\rm{d,c0}}}
\newcommand{\Bzero}{B_0}
\newcommand{\Bzc}{B_{z,\rm{c}}}
\newcommand{\Bzcb}{B_{z,\rm{c0}}}
\newcommand{\Beqcb}{B_{z,\rm{eq,c0}}}
\newcommand{\sign}{\sigma_{\rm{n}}}
\newcommand{\siggeo}{\sigma_{\rm{geo}}}
\newcommand{\Fmag}{F_{\rm{mag},r}}
\newcommand{\vn}{v_{\rm{n}}}
\newcommand{\vnu}{v_{\rm{n,u}}}
\newcommand{\vi}{v_{\rm{i}}}
\newcommand{\vnms}{v_{\rm{ms}}}
\newcommand{\van}{v_{\rm{A}}}
\newcommand{\vanu}{v_{\rm{A,u}}}
\newcommand{\Alf}{Alfv\'{e}n}
\newcommand{\signc}{\sigma_{\rm{n,c}}}
\newcommand{\rhon}{\rho_{\rm n}}
\newcommand{\rhonc}{\rho_{\rm{n,c}}}
\newcommand{\kB}{k_{{}_{\rm B}}}
\newcommand{\kc}{k_{\rm c}}
\newcommand{\zcr}{\zeta_{{}_{\rm{CR}}}}
\newcommand{\mcent}{M_{\rm{cent}}}
\newcommand{\mcentdot}{\dot{M}_{\rm{cent}}}
\newcommand{\phicent}{\Phi_{B,\rm{cent}}}
\newcommand{\phicentdot}{\dot{\Phi}_{B,\rm{cent}}}
\newcommand{\mdot}{\dot{M}}
\newcommand{\phidot}{\dot{\Phi_{\rm B}}}
\newcommand{\rp}{r^{\prime}}
\newcommand{\rinner}{r_{\rm{inner}}}
\newcommand{\tint}{t_{\rm{int}}}
\newcommand{\aB}{a_B}
\newcommand{\gaminst}{\gamma_{{}_{\rm{II}}}}
\newcommand{\Pext}{P_{\rm{ext}}}
\newcommand{\Pextd}{\tilde{P}_{\rm{ext}}}
\newcommand{\lref}{\tilde{l}_{\rm{ref}}}
\newcommand{\delt}{\Delta t}
\newcommand{\delshk}{\Delta_{\rm{shk}}}
\newcommand{\tni}{\tau_{\rm{ni}}}
\newcommand{\tin}{\tau_{\rm{in}}}
\newcommand{\tad}{\tau_{{}_{\rm{AD}}}}
\newcommand{\tacc}{\tau_{\rm{acc}}}
\newcommand{\ts}{\tau_{\rm s}}
\newcommand{\tgr}{\tau_{\rm{gr}}}
\newcommand{\tff}{\tau_{{}_{\rm{ff}}}}
\newcommand{\vd}{v_{{}_{\rm D}}}
\newcommand{\vff}{v_{{}_{\rm{ff}}}}
\newcommand{\vdcrit}{v_{{}_{\rm{D,crit}}}}
\newcommand{\gr}{g_r}
\newcommand{\ssig}{s_\sigma}
\newcommand{\sdelt}{s_{{}_{\delt}}}
\newcommand{\sB}{s_B}
\newcommand{\fsig}{\cal{F}_\sigma}
\newcommand{\an}{a_{\rm n}}
\newcommand{\sigw}{\langle \sigma w \rangle_{\rm{ni}}}
\newcommand{\sigHI}{\langle \sigma w \rangle_{\rm{i \HII}}}
\newcommand{\sigHII}{\langle \widetilde{\sigma w} \rangle_{\rm{i \HII}}}
\newcommand{\siggna}{\langle \sigma w \rangle_{\rm{gn}}}
\newcommand{\siggn}{\langle \widetilde{\sigma w} \rangle_{\rm{gn}}}
\newcommand{\rzero}{R_0}
\newcommand{\rcrit}{R_{\rm{crit}}(t_j)}
\newcommand{\rcritb}{R_{\rm{crit}}(\delt_j)}
\newcommand{\lamTcrit}{\lambda_{\rm{T,crit}}(t_j)}
\newcommand{\lamTcritb}{\lambda_{\rm{T,crit}}}
\newcommand{\rfront}{r_{{}_{\rm{HMD}}}}
\newcommand{\vfront}{v_{{}_{\rm{HMD}}}}
\newcommand{\czero}{\rm{C_0}}
\newcommand{\mgzero}{\rm{Mg_0}}
\newcommand{\nazero}{\rm{Na_0}}
\newcommand{\fezero}{\rm{Fe_0}}
\newcommand{\szero}{\rm{S_0}}
\newcommand{\sizero}{\rm{Si_0}}
\newcommand{\zC}{\zeta_{{}_{\czero,\rm{UV,CR}}}}
\newcommand{\zMg}{\zeta_{{}_{\mgzero,\rm{UV,CR}}}}
\newcommand{\zNa}{\zeta_{{}_{\nazero,\rm{UV,CR}}}}
\newcommand{\zFe}{\zeta_{{}_{\fezero,\rm{UV,CR}}}}
\newcommand{\zS}{\zeta_{{}_{\szero,\rm{UV,CR}}}}
\newcommand{\zSi}{\zeta_{{}_{\sizero,\rm{UV,CR}}}}
%
\title{DYNAMICAL COLLAPSE OF NONROTATING MAGNETIC MOLECULAR CLOUD CORES: EVOLUTION THROUGH POINT-MASS FORMATION}
\author{Glenn E. Ciolek and Arieh K\"{o}nigl}
\affil{Department of Astronomy and Astrophysics,
University of Chicago \\
5640 S. Ellis Avenue, Chicago, IL 60637 \\
\vspace{2ex}
Accepted for publication in {\it The Astrophysical Journal}, 1 Sept 1998 issue}
\begin{abstract}
\vspace{-3ex}
We present a numerical simulation of the
dynamical collapse of a nonrotating, magnetic molecular cloud
core and follow the core's evolution through the formation of a central
point mass and its subsequent growth to a $1~\msol$
protostar. The epoch of point-mass formation  (PMF)
is investigated by a self-consistent extension of previously presented
models of core formation and contraction in axisymmetric,
self-gravitating, isothermal, magnetically supported interstellar
molecular clouds. Prior to PMF, the core
is dynamically contracting and is not well approximated by a
quasistatic equilibrium model. Ambipolar diffusion, which plays
a key role in the early evolution of the core, is
unimportant during the dynamical pre-PMF collapse
phase. However, the appearance of a
central mass, through its effect on the gravitational field in
the inner core regions, leads to a ``revitalization'' of
ambipolar diffusion in the weakly ionized
gas surrounding the central protostar. This
process is so efficient that it leads to a decoupling of the
field from the matter and results in an outward-propagating
hydromagnetic C-type shock. The existence of an ambipolar
diffusion-mediated shock of this type was predicted
by Li \& McKee (1996), and we find that the basic shock
structure given by their analytic model is well reproduced
by our more accurate numerical results. Our calculation also
demonstrates that ambipolar diffusion, rather than Ohmic
diffusivity operating in the innermost core region, is the main field
decoupling mechanism responsible for driving the shock after PMF.

The passage of the shock leads to a 
substantial redistribution, by ambipolar diffusion but possibly
also by magnetic interchange, of the mass contained
within the magnetic flux tubes in the inner core.
In particular, ambipolar diffusion reduces the flux initially
threading a collapsing $\sim 1~\msol$ core by a factor $\simgt
10^3$ by the time this mass accumulates within the inner radius
($\simeq 7.3~{\rm AU}$) of our computational grid. This
reduction, which occurs primarily during the post-PMF phase of
the collapse, represents a significant step towards the
resolution of the protostellar magnetic flux problem.

Our calculations indicate that a $1~\msol$ protostar forms in
$\sim 1.5 \times 10^5~\rm{yr}$ for typical cloud parameters.
The mass accretion rate is time dependent, in part because of
the C-shock that decelerates the infalling matter as it
propagates outward: the accretion rate rises to
$\simeq 9.4~\msol~\rm{Myr}^{-1}$ early on and decreases to
$\simeq 5.6~\msol~{\rm{Myr}}^{-1}$ by the time a solar-mass protostar
is formed. The infalling gas disk surrounding the protostar has a mass
$\sim 10^{-2}~\msol$ at radii $r \simgt 500~\rm{AU}$. A distinguishing
prediction of our model is that the rapid ambipolar diffusion after the
formation of a protostar should give rise to large
($\simgt 1~{\rm{km}}~{\rm s}^{-1}$), and potentially measurable,
ion--neutral drift speeds on scales $r \simlt 200~\rm{AU}$.

The main features of our simulation, including the C-shock
formation after PMF, are captured by a similarity solution that
incorporates the effects of ambipolar diffusion (Contopoulos, Ciolek, \&
K\"{o}nigl 1997).
\end{abstract}
\vspace{-3ex}
\section{Introduction}
It is generally accepted that most of the star-formation activity in our
galaxy takes place through the gravitational collapse of molecular
cloud cores (e.g., Mouschovias 1987; Shu, Adams, \& Lizano 1987). It is,
furthermore, believed that interstellar magnetic fields play
a central role in this process in that their stresses are the
dominant agent that acts against gravity to prevent, or delay,
cloud collapse (e.g., Mouschovias 1978; McKee et al. 1993). This is
embodied in the concept of a critical mass $M_{\rm crit}$, which in
general takes account of both the magnetic and the thermal pressure
contributions to the support of the cloud, but
which, in the case that magnetic stresses dominate, reduces to
$M_{\rm crit} \approx 0.13 \phiB/G^{1/2}$, where $G$ is the
gravitational constant and $\phiB$ is the magnetic flux that threads the
cloud. Clouds whose mass $M$ exceeds
$M_{\rm crit}$ are ``supercritical'': they collapse on the free-fall
timescale. In contrast, ``subcritical'' clouds ($M < M_{\rm crit}$) can
avoid collapse on
the much longer ambipolar diffusion timescale. In the latter
case, the neutrals gradually contract by diffusing inward
through the ions and field, leaving behind a magnetically supported
envelope and eventually forming a supercritical core that
undergoes dynamical collapse (e.g., Mouschovias 1996).

Because of the complexity of the problem --- it involves solving
the full dynamical equations of a magnetized, multicomponent (neutrals,
ions, electrons, as well as charged and neutral grains) fluid
that evolves over many decades in size and density in a
nonspherically symmetric manner (because of the presence of
ordered magnetic fields and likely also rotation) --- much of the
progress in this area has been accomplished through the use of
numerical simulations. One of the main efforts to simulate core
formation and contraction due to ambipolar diffusion in
magnetically supported molecular clouds has been carried out by
Mouschovias and coworkers (e.g., Fiedler \& Mouschovias 1992,
1993; Ciolek \& Mouschovias 1993, 1994, 1995, hereafter CM93, CM94,
CM95, respectively; Basu \& Mouschovias 1994, 1995a, b). These
studies followed the evolution of the core over six decades in
density up to central densities $\sim 3 \times 10^9~\cc$, where
the assumption of isothermality starts to break down because of
radiative trapping (e.g., Gaustad 1963; Hayashi 1966). This assumption had been
adopted in the interest of simplicity: sophisticated and computationally
intensive numerical techniques are generally needed to calculate
the thermal structure of the gas during the opaque phase of protostellar
evolution (e.g., Larson 1969, 1972; Winkler \& Newman 1980; Stahler,
Shu, \& Taam 1981; Boss 1984;
Myhill \& Kaula 1992; Myhill \& Boss 1993). As a result of this
restriction, the aforementioned calculations did not follow the
collapse of the core to the time where a point mass -- a protostar
-- is formed at the center, although they did obtain
valuable information on the conditions leading to this critical
event. In particular, by the time these simulations were
terminated, the inner region of the core was collapsing
dynamically and was characterized by neutral infall speeds $\sim
C$ (the isothermal speed of sound) and inward accelerations $\simgt 0.3 |\gr|$
[where $\gr(r)$ is the local gravitational
acceleration]. Furthermore, the thermal pressure, while remaining
relatively unimportant in the envelope, came to exceed the
magnetic pressure near the center. Basu (1997)
derived a time-dependent, semianalytic solution that extended these
ambipolar diffusion models up to the instant of point-mass
formation (hereafter referred to as PMF \footnote{Creation
of a central point mass was commonly referred to in previous
papers as ``core formation.'' In this paper we use
the term ``point-mass formation'' so as not to confuse this process
with the formation of an extended, magnetically supercritical,
molecular cloud core.}). He found that ambipolar diffusion
continues to gradually erode the retarding magnetic forces in
the inner core, making the collapse increasingly more dynamical
(and the thermal-to-magnetic pressure ratio in the inner core
progressively larger) as PMF is approached.

The diminution of magnetic forces in the innermost regions of a
collapsing core just prior to PMF suggests that one could gain
some insight into the protostar formation process from previous
studies of PMF in {\it nonmagnetic}, spherically symmetric,
isothermal clouds.  Analytic similarity solutions have uncovered two limiting
behaviors: Penston (1969) and Larson (1969) found a solution in
which, just before PMF, the infall speed approaches $\sim 3.3~C$
at all radii while the density scales with radius as $r^{-2}$,
resulting in a spatially uniform mass inflow rate
$\sim 29~C^3/G$ (where $G$ is the gravitational
constant). Hunter (1977) extended this solution
past PMF and showed that, immediately after the central mass
is formed, the accretion rate onto the protostar increases to
$\sim 47~C^3/G$. In the other limit, Shu (1977) obtained a
solution that is static prior to PMF (with the density distribution of a
singular isothermal sphere, which also scales as $r^{-2}$) and
that takes on an expansion-wave character (with a constant mass
accretion rate $\sim 0.98~C^3/G$ onto the protostar) following
PMF. Hunter (1977) and Whitworth \& Summers (1985) demonstrated
that there are, in fact, infinitely many similarity solutions
that span the range between the Larson-Penston and Shu results,
with the nature of any given solution being determined by the
initial configuration of the cloud and the conditions at its
boundary. Numerical simulations carried out by Hunter (1977) and
by Foster \& Chevalier (1993) confirmed the dependence on the
initial and boundary conditions. In particular, it was found
that the behavior of the central regions of clouds that are
initially marginally stable to collapse approximates that of the
Larson-Penston solution at the PMF epoch, although it was
determined that the mass accretion rate onto the protostar declines at
later times. It was, however, also found that the post-PMF
evolution of clouds that initially have more extended envelopes
approximates that of the Shu solution at late times. Since the
initial and boundary conditions of real clouds are expected to
depend on the detailed configuration and evolution of the embedded
magnetic field, it is clear that one needs to incorporate
magnetic field effects into the collapse calculations to
adequately model the formation of protostars.

There have been several recent attempts to calculate PMF following the
collapse of magnetic interstellar clouds. Although they have all
contributed to our understanding of the processes involved,
their results were hampered by the adopted  assumptions or
approximations. For example, Tomisaka (1996) modeled clouds that
had equal thermal and magnetic energy densities, so that they
were not primarily supported by magnetic fields. This means that
his model clouds were magnetically supercritical. This
assumption is at variance with H {\sc I} and OH Zeeman
measurements of magnetic field strengths in molecular clouds, which are
consistent with models of magnetically subcritical clouds (Crutcher et
al. 1993, 1994, 1996). Li \& Shu (1997) modeled
PMF in self-similar, magnetic cores. They
assumed that cores immediately before PMF can be represented by
hydrostatic configurations of singular isothermal disks
and that the magnetic flux is
frozen into the neutrals. These assumptions are inconsistent
with the above-cited results of numerical simulations
and semianalytic solutions of the
collapse of magnetically supported molecular clouds that undergo
ambipolar diffusion (as well as with the simulations of
thermally supported spherical clouds), which have found that
the inner core regions collapse dynamically as PMF is
approached (see also \S 3.2). As we show below, ambipolar
diffusion, which plays a
key role in bringing about the dynamical collapse, is generally
important also {\it after} PMF. Safier, McKee, \& Stahler
(1997) did include ambipolar diffusion in the modeling of
magnetic cloud collapse and post-PMF evolution. However, their
model is strictly spherical, and they
neglected the effect of thermal pressure gradients in comparison
with magnetic stresses. Furthermore, their formulation is quasistatic
and does not incorporate the magnetic induction equation for the time
evolution of the magnetic field. Li (1998) extended the Safier et al.
model by including thermal-pressure and time-dependent terms and by
adding the induction equation. This enabled him to follow the time
evolution of his model cores (for $r > 150~\rm{AU}$) even during the
dynamical phases of the collapse. However, by retaining the
spherical-symmetry assumption of Safier et al., his calculations were
also unable to yield the geometry of the magnetic field lines.
\footnote{The same is true for
the spherically symmetric self-similar model of a
collapsing magnetic cloud devised by Chiueh \& Chou (1994),
which, however, does not include ambipolar diffusion. It should
be noted that all models that assume spherical symmetry
do not satisfy the solenoidal condition
$\nabla \cdot \Bvec =0$ on the magnetic field.}

The importance of ambipolar diffusion after PMF can be inferred by comparing
the ambipolar diffusion timescale $\tad = r/\vd$ (where $\vd$
is the ion--neutral drift speed) and the gravitational contraction
($\simeq$ free-fall) timescale $\tgr = (r/|\gr|)^{1/2}$
before and after PMF. In axisymmetric geometry,
$\tad/\tgr \simeq (\tgr/\tni) \, {\mu_ B}^2$ in the
inner flux tubes of a supercritical core (e.g., Mouschovias 1991),
where ${\mu_B}(r)$ is the total mass-to-flux ratio at radius $r$ (in
units of the critical value for gravitational collapse) and
\begin{equation}
\label{tnieq}
\tni = 1.4 \left[1 + 0.067 \frac{\left(\mHII /2~\rm{a.m.u.}\right)}{\left(\mi/30~\rm{a.m.u.}\right)}\right] \frac{1}{\nni \sigw}
\end{equation} 
is the neutral--ion collision time. In equation (\ref{tnieq}),
$\nni$ is the ion density and $\sigw$ is the average collisional
rate between ions of mass $\mi$ and neutral $\HII$ molecules of
mass $\mHII$ ($\simeq 1.7 \times 10^{-9}~{\rm{cm^3}}~{\rm s}^{-1}$
for collisions between neutrals and $\Mgp$ or $\HCOp$ ions; McDaniel
\& Mason 1973); the factor 1.4 accounts for a 20\% helium abundance
by number. CM94 and CM95 found that, to first order for the
late pre-PMF evolution of cores in disk-like clouds, the magnetic field
$B \approx 3\, (r_0/r)~\rm{mG}$ (where $r_0 = 40~\rm{AU}$),
$\nni \approx 0.1~\cc$ (valid for neutral densities
$\nn \simgt 10^7~\cc$; see Figs. $2b$ and $4b$ in CM94), the vertical
column density $\sign \approx 5\, (r_0/r)~{\rm g}~\rm{cm}^{-2}$, and
the total mass $M(r) \approx 6 \times 10^{-3} (r/r_0) \msol$. 
For a disk-like cloud, the critical
mass-to-flux ratio $(M/\phiB)_{\rm{d,crit}}= (4 \pi^2 G)^{-1/2}$,
where $G$ is the gravitational constant (Nakano \& Nakamura 1978).
Therefore, one finds $\mu_B \approx 2.7$, $|\gr| \approx G
M(r)/r^2 \approx 2 \times 10^{-6} (r_0/r) ~{\rm{cm}}~\rm{s}^{-2}$,
and $\tni \approx 7 \times 10^9~\rm{s}$, which yields $\tgr
\approx 2 \times 10^{10} (r/r_0) ~\rm{s}$
and $\tad/\tgr \approx 20\, (r/r_0)$. It follows that
$\tad/\tgr \gg 1$ for $r \gg 2~\rm{AU}$. Hence, as has already been
demonstrated in the past, ambipolar diffusion is ineffective as a
dynamically collapsing core approaches PMF, and for $r \gg 2~\rm{AU}$
the magnetic flux can be considered frozen into the neutrals. (For
reasons discussed in \S 2.2, we do not consider $r \simlt 5~\rm{AU}$.)
Turning now to the post-PMF epoch, when the central point mass
comes to dominate the gravitational field in the
innermost flux tubes, we use the relation
$\tad/\tgr \approx \left(1- |\an|/|\gr|\right)^{-1} (\tgr/\tni)$,
where $\an$ is the inward acceleration of the neutrals and $\gr$
is the total gravitational acceleration (see
eqs. [\ref{rforceeq}] and [\ref{tadeq}] in \S 3.3).
In this case, normalizing $r$ as before,
$|\gr| \approx 4 \times 10^{-4} (\mcent/\msol) (r_0/r)^2
{\rm{cm}}~\rm{s}^{-2}$.
Substituting again $\tni \approx 7 \times 10^9~{\rm s}$
(corresponding to $\nni \approx 0.1~\cc$) , the above ratio becomes
$\tad/\tgr \approx 0.2\,
(r/r_0)^{3/2}(\msol/\mcent)^{1/2}\left(1-|\an|/|\gr|\right)^{-1}$.
For $|\an|/|\gr|$ in the range 0.2 -- 0.9, which corresponds to
the period after PMF when the collapse becomes progressively
more dynamical, one infers $\tad/\tgr \approx (0.2 - 2)
(r/r_0)^{3/2}(\msol/\mcent)^{1/2}$. Hence
$\tad \simlt \tgr$ for $\mcent \simgt 0.1 \msol$ and
$r \simlt 50~\rm{AU}$. [A similar result is obtained if one
continues to use the pre-PMF relations and simply substitutes $\mcent$
for $M(r)$.]
This estimate indicates that {\em decoupling of the gas and
magnetic field by ambipolar diffusion should occur in the
inner core regions after PMF}. The physical reason for this is that
the strength of the gravitational field in the weakly
ionized gas near the origin is greatly enhanced by the
appearance and growth of a central point mass, causing the
neutrals to fall in more rapidly while the plasma and magnetic
field are left behind. (The same basic reason --- the appearance
of a progressively growing free-fall zone around the origin
following PMF --- is also the cause of the increase in the mass
inflow rate into the center at that epoch first discovered in
the above-referenced nonmagnetic collapse calculations.) The
foregoing conclusion is verified by a detailed calculation in
\S~3.3, where we show that ambipolar diffusion after PMF is, in
fact, so efficient that it
effectively decouples the neutrals and magnetic field in the
innermost core region, with dramatic consequences for the
subsequent dynamical evolution of the core. An alternative, yet
equivalent, analysis of the effectiveness of ambipolar diffusion
following PMF, based on the scaling of magnetic forces (particularly
the magnetic tension force) after PMF, is given in Appendix C.

In this paper we present a
detailed numerical simulation of point-mass formation in a nonrotating,
magnetic, dynamically collapsing protostellar core, properly
accounting for the effect of ambipolar diffusion. Unlike earlier
studies, we use an initial state that is consistent with the realistic
models of core formation and collapse, as presented earlier by Mouschovias
and coworkers. As we noted above, the simulations carried out by
that group were terminated at densities where the isothermality
assumption started to become invalid because of radiative trapping.
Although a proper treatment of radiative trapping is
indispensable for a complete treatment of the star-formation
process, one can adopt a simpler approach that circumvents this difficulty by
removing the region of radiative trapping from the active computation mesh. 
This is justified by the fact that the region of radiative
trapping (typically a few AU) is several orders of magnitude
smaller than the characteristic core size ($r_{\rm{core}}
\approx 0.1~\rm{pc}$). This region can therefore be considered
to be effectively point-like, and one can proceed to calculate
the formation of a central point mass within a gravitationally
collapsing core and its effect on the subsequent evolution of
the core by retaining the isothermality assumption. In adopting
this approach, we note that the assumption
of isothermality was also employed in previous studies of
protostar formation in nonmagnetic cloud cores as well as in the
more recent attempts to model PMF in magnetic clouds.

The plan of the paper is as follows. In \S~2 we review the main
characteristics of the pre-PMF core-evolution calculations of
CM93, CM94, and CM95 and outline the model modifications that
we have implemented to extend the simulations beyond PMF. In
\S~3 we present the results of our calculations. We consider a
typical model and follow the evolution of all physical quantities of
interest through the PMF epoch. We also describe the formation (due to
ambipolar diffusion in the innermost core after PMF) of a
C-type shock and consider its propagation through the collapsing
core. The appearance of such a shock as a result of
field--matter decoupling was first pointed out by Li
\& McKee (1996), who proposed that the relevant field decoupling
mechanism was Ohmic dissipation in the
innermost regions of the core. Our study, however, reveals that
ambipolar diffusion occurring outside the region
of Ohmic dissipation is the main cause of magnetic flux
decoupling after PMF. In \S~4 we discuss the structure of the
ambipolar diffusion-mediated shock and show that our numerical
calculations qualitatively reproduce the predictions of the simplified
analytic model constructed by Li \& McKee (1996). We
present a quantitative comparison with their results and also
address the issue of interchange instability in the post-shock
region, first raised in their work, in light of our detailed
computations. In that section we also discuss the observational
implications of our simulation and briefly comment on the
magnetic flux problem during star formation. Our results are summarized
in \S~5.
\section{Characteristics of Model Clouds}
\subsection{Pre-PMF Phase}
For the pre-PMF phase of core evolution, which encompasses the formation
(due to ambipolar diffusion) and contraction of magnetically supercritical
cores in magnetically subcritical clouds, we use the models described
in CM93, CM94, and CM95. Further details are given in Appendix A.
Here we summarize key features of these models.

Clouds are modeled as axisymmetric thin disks, with half-thickness $Z(r,t)$,
embedded in a constant-pressure external medium, with the axis of symmetry
aligned with the $z$ axis of a cylindrical polar coordinate
system $(r,\phi,z)$;
exact balance between thermal-pressure and gravitational forces along
magnetic field lines is assumed to hold at all times.
These simplifying assumptions are based on the results of
Fiedler \& Mouschovias (1992, 1993), who found that initially uniform,
spherical or cylindrically symmetric, self-gravitating magnetic
molecular clouds rapidly flatten and establish force balance (between
thermal-pressure and gravitational forces) along magnetic field lines.
In their typical models, balance of forces along field lines 
was maintained even after the onset of dynamical contraction
perpendicular to the field lines. The collisional friction of (charged and
neutral) grains, which in certain cases can be significant (CM94, CM95),
is ignored in this paper. We also neglect the effect of
rotation and magnetic braking (e.g., Basu \& Mouschovias 1994, 1995a, b);
they will be considered in a later paper.

Model clouds are initially in magnetohydrostatic equilibrium and would
remain so indefinitely
in the absence of ambipolar diffusion. Hence, any evolution is entirely
the result of ambipolar diffusion, which is ``turned on'' at time $t=0$. 
The evolution of a magnetically supercritical core is followed up
to a time $\tint$, at which the central density exceeds $10^{11}~\cc$.
At this point the pre-PMF calculation is halted. We then proceed
as described in the next subsection.
\subsection{PMF Phase and Subsequent Evolution; Modification of the
Ambipolar Diffusion Models}
To calculate this phase of the evolution we use as initial data the
values of the physical quantities (column density, magnetic field,
neutral infall speed, etc.) of the pre-PMF epoch at time $\tint$.
To numerically calculate through the PMF phase of evolution, we use the
``central sink cell'' method originally employed by Boss \& Black
(1982) to model the collapse of nonmagnetic, isothermal
clouds. This method was also used in the collapse studies of Foster
\& Chevalier (1993) and Tomisaka (1996). In this method we make the central cell of
our (stationary) computational mesh a sink cell, with size equal
to the mesh inner boundary
radius $r = \rinner$: the central sink cell is essentially a ``hole''
at the origin of our computational mesh. Use of the central sink cell
keeps the numerical time step $\delta t_{\rm{num}}$ [$\sim \rinner/|\vn(\rinner)|$,
where $\vn$ is the infall speed of the neutrals] finite throughout the
PMF epoch. 

We fix the value of $\rinner$ at the radius at which the governing
equations of a model cloud no longer remain valid. Two fundamental
assumptions of our models are isothermality and the freezing of
magnetic flux into the ions. The isothermal approximation breaks down
for $\nn \simgt 10^{11}~\cc$ (Gaustad 1963; Hayashi 1966; Larson 1969;
Winkler \& Neuman 1980); at densities above this value, radiative
heating of the gas can no longer be neglected. Freezing of the magnetic
field in the ions is no longer valid when the dimensionless ion magnetic attachment
parameter $\Gamma_{\rm i} \equiv \omega_{\rm i} \tin$, also known as
the ion Hall parameter, is $\simlt 1$ (see eqs. [38] and [39] of CM93),
where $\omega_{\rm i}$ is the ion
gyrofrequency and $\tin$ is the ion--neutral collisional timescale.
\footnote{When $\Gamma_{\rm i} < 1$ the assumption of flux freezing
in the ions is no longer valid. However, one can still consider the
magnetic flux to be frozen into the electrons, if the electrons
are the main charge carriers (as opposed to charged grains) and
the electron magnetic attachment parameter $\Gamma_{\rm e} \equiv \omega_{\rm e} \tau_{\rm{en}} > 1$
(for a discussion, see the Appendix of K\"{o}nigl 1989). At even higher
densities than we consider here, $\Gamma_{\rm e}$ becomes $< 1$ and the
effect of Ohmic dissipation has to be
included in the magnetic induction equation
(Spitzer 1963; Pneuman \& Mitchell 1965; Norman
\& Heyvaerts 1985; Nakano \& Umebayashi 1986a,b). For detailed
discussions of the
effect of a finite conductivity on the MHD equations for multicomponent
plasmas at the high densities encountered in collapsing cores and/or
protostellar accretion disks we refer the reader to Nakano (1984),
Draine (1986), K\"{o}nigl (1989), Wardle \& K\"{o}nigl (1993),
Mouschovias (1996), and Wardle \& Ng (1998).}
The CM94 collapse model that neglects the collisional effect of grains
(see their Figs. $2a$ and $2f$) yields the approximate
scaling $\Beq \simeq 3.5 \times 10^3 (\nn/2.6 \times 10^9~\cc)^{0.4}
\mu {\rm G}$ for the equatorial ($z=0$) magnetic field. 
Inserting this relation into equation (40) of CM93 gives
$\Gamma_{\rm i} \simlt 1$ for $\nn \simgt 5 \times 10^{10}~\cc$. 
Thus, coincidentally, the assumption of freezing of magnetic
flux in the ions breaks
down at effectively the same density as the one where the isothermality approximation also becomes
invalid. We take $\rinner$ to be the radius at which 
$\nn(\rinner) \simeq 10^{11}~\cc$. Typically, this
corresponds to $\rinner \simgt 5~\rm{AU}$ (see \S~3.3).

Mass and magnetic flux accumulate in the central sink
from our active computational region, consisting of all the cells located
at $r \geq \rinner$; the rate at which mass is advected into the central
sink is
\begin{equation}
\label{dmdtcenteq}
\mcentdot \equiv
\frac{\partial \mcent}{\partial t} =
\left. -2 \pi r \sign \vn \right|_{r=\rinner},
\end{equation}
and the rate at which magnetic flux flows into the central cell is
\begin{equation}
\label{dphidtcenteq}
\phicentdot \equiv \frac{\partial \phicent}{\partial t} =
\left. -2 \pi r \Beq \vi \right|_{r=\rinner},
\end{equation}
where $\mcent(t)$ and $\phicent(t)$ are the central mass and magnetic
flux, $\sign(r,t)$ and $\Beq(r,t)$ are the column density and magnetic
field strength (in the equatorial plane of the disk), and $\vn(r,t)$ and
$\vi(r,t)$ are the neutral and ion infall speeds.

Despite the fact that the mass and flux contained within the central sink cell
are exterior to the active computational region in our models, they
still affect the evolution of the collapse through their gravitational
and magnetic influence on the matter at $r \geq \rinner$. This has to be
incorporated into the calculation of the radial gravitational
and magnetic field components. The radial gravitational acceleration
in a thin disk is
(see CM93, eqs. [31a] and [31b])
\begin{equation}
\label{oldgreq}
\gr(r) =
2 \pi G \frac{d}{dr}
\int^{\infty}_{0}
d \rp~\rp~\sign(\rp) \int^{\infty}_{0} dk~J_0(kr) J_0(k \rp)
= - 2 \pi G \int^{\infty}_{0} dk~J_1(k r) \int^{\infty}_{0} d \rp~\rp \sign(\rp) J_0(k \rp), 
\end{equation}
where $J_0$ and $J_1$ are
Bessel's functions of order 0 and 1, respectively. One can show
by direct integration of equation (\ref{oldgreq}) that, for $\sign(r) \propto r^{\ssig}$,
where $\ssig$ is a constant between -1/2 (corresponding to free-fall
during the post-PMF phase; see \S~3.3) and -1 (typical of the
late stages of the pre-PMF phase, see \S~3.2), $\gr(r) =  -{\fsig} M(r)/r^2$, where
$M(r)$ is the mass enclosed within the radius $r$ and ${\fsig}$ is a constant
between 0.68 and 1. As an example, column density profiles $\sign(r) \propto r^{-0.95}$ 
typically developed in the 
inner regions of the contracting cores of the models presented 
by CM94 and CM95 (see also \S~3.2), corresponding to 
${\fsig} \simeq 0.82$ in the above expression. Considering
the alternative case of a highly localized source, such as
a uniform-density sphere of radius $R_\star$ ($\ll \rinner$)
and mass $M_\star$, the column density obeys the relation
\leteq
\begin{eqnarray}
\sign(r) &=& \frac{3}{2 \pi} \frac{M_\star}{R_\star^2}\left[1 -
\left( \frac{r}{R_\star}\right)^2 \right]^{1/2},  \   r \leq
R_\star \ , \\
&=& 0\ ,    \hspace{10em} r > R_\star \ ,
\end{eqnarray}
\beq
and one finds from the solution of equation (\ref{oldgreq}) that, for
$r > R_\star$, $\gr(r) = -G M_\star/r^2$, as expected.
Hence $\gr(r > \rinner,t) \simeq -G M(r,t)/r^2$
for the cases of a power-law column density
profile and of a localized mass source. As these two extremes tend to
bracket the range of possible behaviors of $\sign(r,t)$ in the innermost
flux tubes of the core, we may use
the expression
\begin{equation}
\label{newgreq}
\gr(r) = -\frac{G \mcent}{r^2} + 2 \pi G \frac{d}{dr} \int_0^\infty d \rp~\rp~\sign(\rp) \int_0^\infty dk~J_0(k r) J_0(k \rp)
\end{equation}
during the PMF epoch. Extension of the integral in equation
(\ref{newgreq}) 
to $r=0$ is obtained, with negligible error, by setting $\sign( r < \rinner) = \sigma_0$,
where $\sigma_0$ is a constant that satisfies the condition $0 \leq \sigma_{0} \leq \sign(\rinner)$. A description of the numerical method used to solve
the integral term on the r.h.s. of equation (\ref{newgreq}) is given
in \S~2.4 of Morton, Mouschovias, \& Ciolek (1994, hereafter MMC94).

As shown in \S~3.1.1 of CM93, the $r$ component of the magnetic field
at the surface of the disk, which appears in the restoring
magnetic tension force (see CM93, eq. [28c]), is given by
\begin{eqnarray}
\label{oldbreq}
\BrZ(r) &=& - \frac{d}{dr} \int^{\infty}_{0} d \rp~\rp \left[\Beq(\rp) - \Bref \right]
\int^{\infty}_{0} dk~J_0(kr) J_0(k \rp) \nonumber \\ 
&=& \int^{\infty}_{0} dk~J_1(k r) 
\int^{\infty}_{0} d \rp \rp \left[\Beq(\rp)-\Bref
\right]~J_{0}(k \rp) \ ,
\end{eqnarray}
where the constant $\Bref$ is the external (background) magnetic
field strength at $r,z \rightarrow \infty$. Typically, $\Beq(r) \gg \Bref$
in the inner flux tubes of a contracting core (CM94, CM95; Basu \& Mouschovias 1994, 1995a, b).
\footnote{Note that, under the conditions $\Beq(r) \gg \Bref$ and
$\Beq(r) \propto \sign(r)$, which in our model calculations
are satisfied before PMF for $ r\ll r_{\rm{core}}$, equations
(\ref{oldgreq}) and (\ref{oldbreq}) yield $\BrZ(r) \propto \gr(r)$.
The condition $\Beq(r) \propto \sign(r)$ expresses flux freezing
into the neutrals, and the relation $\BrZ \propto \gr$ was
derived under this assumption by Li \& Shu (1997) and Zweibel \&
Lovelace (1997), who considered model clouds with $\Bref=0$.
This relation was also employed by Basu (1997) in modeling
pre-PMF core collapse.}

By analogy with the discussion of the gravitational field $\gr(r)$ in the
preceding paragraph, calculation of $\BrZ(r)$ for the bracketing
source magnetic field profiles $\Beq(r) \propto r^{-\sB}$, and that of a
central point-flux source [i.e., $\Beq(r)=B_{\star}=const$
for $r \leq R_\star$, and $\Beq(r) = 0$ for $r > R_\star$], 
equation (\ref{oldbreq}) yields, for both cases,
$\BrZ(r) \simeq \Phi_{B}(r)/2 \pi r^2$, where $\Phi_{B}(r)$ is the
total magnetic flux enclosed within radius $r$. It follows then, using
a derivation similar to that of equation (\ref{newgreq}), that
$\BrZ(r > \rinner,t)$ after PMF can be written as
\begin{equation}
\label{newbreq}
\BrZ(r) = \frac{\phicent}{2 \pi r^2} - \frac{d}{dr} \int^{\infty}_{0} d \rp~\rp \left[\Beq(\rp) - \Bref \right]
\int^{\infty}_{0} d k~ J_0(k r) J_0(k \rp) \ .
\end{equation}
Extension of the integral on the r.h.s. of equation (\ref{newbreq})
to $r < \rinner$ is done in a fashion similar to that discussed above
in connection with equation (\ref{newgreq}).

The effect of the gravitational field of the central point mass also
needs to be included in the equation of quasistatic equilibrium along
magnetic flux tubes. Assuming balance of gravitational and thermal-pressure
forces in the $z$ direction
within the disk yields
\begin{equation}
\label{newforcbaleq}
\rhon C^2 = \Pext + \frac {\BrZ^2}{8\pi} \left [ 1 + \left (
\frac{\partial Z}{\partial r} \right ) ^2 \right ] + \frac{\pi}{2} G \sign^2 +
\frac{G \mcent \rhon}{r}
\left\{1 - \left[1 + \left(\frac{\sign}{2 \rhon r}\right)^2
\right]^{-1/2}\right\}\ .
\end{equation}
In deriving equation (\ref{newforcbaleq}) we have used, as in CM93--CM95,
the ``one-zone approximation'' for the relation between the mass
density $\rhon$ and the column
density $\sign$: $\sign(r,t) = \int_{-Z(r,t)}^{Z(r,t)} \rhon(r,t) dz
= 2 \rhon(r,t) Z(r,t)$. The first three terms on the r.h.s.
of equation (\ref{newforcbaleq}), present even in the absence of a central
point mass (see CM93, eq. [26]), are, respectively, the constant external
pressure, the magnetic squeezing associated with the
radial field component at the disk surface, and the
self-gravitational stress
of the matter contained in a flux tube; the last term on the r.h.s.
of equation (\ref{newforcbaleq}) is the
tidal gravitational stress corresponding to the $z$ component of
the gravitational field of the central point mass. Equation
(\ref{newforcbaleq}) simply states
that the pressure in the equatorial plane is equal to the sum of
the external thermal plus magnetic stresses and the total
gravitational stress acting on each
flux tube; this equation is used to calculate $\rhon(r,t)$.
\footnote{For computational convenience, we continue to use the
quasistatic approximation for motions in the $z$ direction even when
there is dynamical collapse in the $r$ direction. To test the validity
of this assumption, we have also run models that did not employ the
quasistatic assumption along flux tubes and
instead allowed for acceleration of the neutrals in the $z$ direction.
In these models there was initially a short-lived
phase of vertical dynamical relaxation and oscillation in the
central flux tubes, followed by rapid reestablishment of balance of forces
along flux tubes, in agreement with equation (\ref{newforcbaleq}).
The overall evolution of these models differed only
slightly from those in which balance of vertical forces was assumed at
all times. However, the computing time needed to run models that
did not assume vertical force equilibrium was typically an order of
magnitude longer than for models that always used the quasistatic
approximation.}

As it turns out, the first two terms on the r.h.s. of equation
(\ref{newforcbaleq}) have a negligible effect on the disk evolution
in our simulations. In particular,
the ratio of $\Pext$ and the self-gravitational stress is
generally $\ll 1$ in the inner flux tubes of the core (see
eq. [A2]). Furthermore, we have verified that, with the exception of the
innermost zones of the active computational grid near the end of
our typical simulation, the magnetic squeezing term is smaller
than the total gravitational stress in equation
(\ref{newforcbaleq}). Since the properties of these zones have
only a small effect on the core structure at that epoch (in
particular, the ambipolar diffusion-driven shock that propagates
through the core has by that time reached much larger radii; see
\S 3.3), we have, for simplicity, omitted this term altogether
in our calculations. As we show in Appendix B, when the core
approaches free-fall collapse under these conditions,
equation (\ref{newforcbaleq}) implies $\rhon \propto r^{-2}$.
This behavior, which differs from the dependence
$\rhon \propto r^{-3/2}$ that characterizes spherical infall
onto a point mass, results from our use of the one-zone approximation in the
relation between the column density and mass density for the thin-disk
cloud model.

A final modification that is required for the post-PMF phase of collapse
is in the ion force equation,
from which the ion--neutral drift speed $\vd= \vi -\vn $
is derived (see CM93, eqs. [50] and [51]).
This has to do with the fact that,
as first pointed out by Mouschovias \& Paleologou (1981), 
the ion--neutral collision rate $\sigw = 1.7 \times 10^{-9} {\rm{cm}}^3 ~{\rm{s}}^{-1} = const$
(used in the ambipolar diffusion models of CM93--CM95), which was
derived by using the well-known Langevin approximation (see, e.g., Gioumousis \& Stevenson 1958; McDaniel \& Mason
1973), is valid only so long as the drift speed satisfies the criterion
$\sigw/|\vd| \geq \siggeo$, where $\siggeo$ is the
{\em geometric} cross section for ion--$\HII$ collisions ($=\pi
[a_{\rm{ion}} + a_{{}_{\HII}]}^2$,
where $a_{\rm{ion}}$ and $a_{{}_{\HII}}$ are the ion and
$\HII$-molecule radii, respectively).
Therefore, for drift speeds $\vd > \vdcrit=\sigw/\sigma_{\rm{geo}}$,
the quantity $\siggeo |\vd|$ must be used for the collision rate,
and instead of using equation (51) of CM93, the appropriate equation
for the drift speed is 
\begin{equation}
\label{newdrifteq}
\vd(r,t) = \left\{1.4 \left[1 + 0.067\frac{(\mHII/2~{\rm{a.m.u.}})}{(\mi/30~{\rm{a.m.u.}})} \right] \frac{\Fmag(r,t)}{\siggeo \nni(r,t)\sign(r,t)} \right\}^{1/2} ,
\end{equation}
where $\Fmag(r,t)$ is the total (radial) magnetic force per
unit area, which includes both the magnetic tension and pressure stresses
(see eq. [28c] of CM93). The factor 1.4 in equation (\ref{newdrifteq})
reflects the fact that we are neglecting the inertial effect of a
20\% He abundance in the neutral--ion collision timescale $\tni$ (see eq. [\ref{tnieq}]).
For the various species of ions we include in our models (see Appendix A)
we adopt a ``generic'' ion radius of 0.8 \AA; this yields a critical
drift speed $\vdcrit=12~\rm{km}~\rm{s}^{-1}$, only slightly
different from the value of $10~\rm{km}~\rm{s}^{-1}$ obtained by
Mouschovias \& Paleologou (1981), who had assumed a plasma consisting
solely of $\Nap$ ions. Hence, when $\vd(r,t) > \vdcrit$, we use
equation (\ref{newdrifteq}); otherwise, equation (51) of CM93 is
used. As we show in \S~3.3, ambipolar
diffusion can continue to be effective even in dynamically
contracting cores, and drift speeds $\vd > \vdcrit$ may
be attained in certain core regions during the post-PMF epoch.

Spatial discretization and
time integration of the equations governing the evolution of a model
cloud are carried out by the methods described in MMC94 --- with the
exception that spatial derivatives in the first computational cell, with
inner boundary $r=\rinner$, are calculated by using one-sided
differences instead of the three-point technique described
in \S~A2.4 of Mouschovias \& Morton (1991). All computations were
performed on an SGI R-4000 Indigo workstation; running in background,
a typical model calculation took $\sim 2$ weeks to form a central
protostar with a mass of $1~\msol$.

As we show in \S~3.3, ambipolar diffusion significantly
alters the evolution of a collapsing core following PMF. To
verify that the results of our typical simulation represent a real
effect that arises from a physical diffusion process and are not an artifact of
numerical diffusivity, we have run a parallel model that corresponds to
flux freezing into the neutrals for all times $\delt \geq 0$.
(This was accomplished simply by setting $\vi = \vn$ in our numerical
code.) The results of this calculation (detailed in \S~3.3)
have revealed no evidence of numerical diffusion arising
from our finite discretization scheme: each neutral fluid element
was found to evolve with
constant mass and magnetic flux throughout the entire simulation,
as would be expected for a fluid with a frozen-in magnetic field.
In \S~3.3 we therefore use the results of the frozen-flux cloud calculation
as an aid in isolating and interpreting the effects of ambipolar
diffusion in our typical model
(see also Appendix C).
We note in this connection that we
have also run other ambipolar-diffusion models with different
mesh sizes and core resolutions and obtained
essentially the same results as in our typical model. This further
confirms the absence of numerical-diffusion effects in our simulations.
In addition, we have tested the effect of varying the assumed
profile of $\Beq$ for $r < \rinner$ in calculating $\BrZ$ from
equation (\ref{newbreq}), and confirmed that our quantitative
results were insensitive to the detailed form of that profile.

\section{Numerical Simulation}
\subsection{Model Parameters}
We present a typical model of the dynamical evolution of a magnetically
supercritical core through the PMF phase. The values of the dimensionless parameters
of this model are listed in Table 1 and Table 2. They are similar to that of model
$\rm{A_{UV}}$ presented in CM95.
The values of the free parameters selected could be used to represent
a molecular cloud with temperature $T=10~{\rm K}$, radius $\rzero = 4.29~\rm{pc}$,
and total mass $M_{\rm d} = 98.3 \msol$. For a gas of $\HII$ with 
a 20\% He abundance the mean mass of a neutral particle is $\mn = 2.33~\rm{a.m.u.}$
and the isothermal speed of sound is $C=(\kB T/\mn)^{1/2}= 0.19~{\rm{km}}~{\rm s}^{-1}$.
The initial central density $\nncb$, column density $\sigcb$, and
magnetic field strength $\Beqcb$ are $2.6 \times 10^3~\cc$, $5.59 \times
10^{-3}~{\rm g}~{\rm{cm}}^{-2}$, and $35.3~\mu{\rm G}$, respectively.
The dimensional
values of the parameters used in the calculation of the equilibrium
abundances of charged particles, such as the abundances of the different
atomic species, cosmic-ray and UV ionization rates, chemical and
charge-transfer reaction rates, etc., are the same as those cited
in \S~3.1 of CM95, with the exception that the probability
${\cal P}_{\rm i}$ of ions sticking onto grains (see CM93,
eqs. [57b] and [57c])
has been changed from 0.9 to 0.99. For the purposes of calculating ion
abundances, we assume a uniform population of grains with radius
$a=3.75 \times 10^{-6}~\rm{cm}$.
\footnote{Although we do not account for grain-neutral friction in this
model, our results would be little
changed if we had instead used grains with radius $a \simgt 7 \times 10^{-6}~\rm{cm}$
and included the effect of grain-neutral collisions,
as grains of this size or larger contribute only marginally to the
total collisional force on the neutrals (see model 2 of Ciolek 1993).}
\subsection{Pre-PMF Phase: Supercritical Core Formation and Contraction}
Models of the self-initiated (due to ambipolar diffusion) formation
and contraction of magnetically supercritical cores have been
presented and  discussed at length by Fiedler \& Mouschovias (1993), CM94 and CM95, and Basu
\& Mouschovias (1994, 1995a, b); we refer the reader to these papers
for more detailed descriptions of the physics of the formation
of protostellar cores in magnetically supported interstellar clouds.
During the pre-PMF phase, the evolution of the core of the typical model
cloud presented here is similar to that of model $\rm{A_{UV}}$ discussed
at length in \S~3.2.1 of CM95; for brevity, we summarize here only 
some of the features of the core in the pre-PMF phase.

Figures $1a$--$1i$ display, respectively, the density (normalized to 
$\nncb$), column density (normalized to $\sigcb$), mass-to-flux
ratio (in units of the critical value for collapse), magnetic field
strength in the equatorial plane of the disk (normalized to $\Beqcb$),
$r$ component of the magnetic field at the surface of the disk
(normalized to $\Beqcb$), infall speed of the neutrals (normalized to
$C$), drift speed (normalized to $C$), mass infall rate
$\mdot$ (in units of $\msol~{\rm{Myr}}^{-1} = 0.65 C^3/G$), and
ratio of local cloud vertical half-thickness $Z$ and radius $r$
as functions of $r/\rzero$, at eleven different times $t_j$. The times
$t_j$ correspond to when the central density $\nnc(t_j) = 10^j \nncb$;
dimensionally, these times are
$t_0 =0$, $t_1 = 7.56~\rm{Myr}$, $t_2 = 9.29~\rm{Myr}$, $t_3=9.56~\rm{Myr}$,
$t_4 = 9.607~\rm{Myr}$, $t_5 = 9.6167~\rm{Myr}$, $t_6 = 9.6169~\rm{Myr}$,
$t_7 = 9.6190~\rm{Myr}$, $t_8 = 9.61977~\rm{Myr}$, $t_9 = 9.61981~\rm{Myr}$,
and $t_{10} = 9.61983~\rm{Myr}$.
An {\it asterisk} on a curve in Figure 1 locates the
instantaneous radius $\rcrit$ inside which
the total-mass to total-flux
ratio is equal to the critical value for collapse, i.e.,
\begin{equation}
\label{rcriteq}
\left(\frac{M}{\Phi_B} \right)_{r=\rcrit} = \frac{1}{ 2 \pi
G^{1/2}}\ .
\end{equation}
An {\it open circle} locates the instantaneous critical thermal
($\simeq$ Jeans) lengthscale $\lamTcrit$ [$=C^2/2 G \signc(t_j)$,
where $\signc(t_j)$ is the central column density at time $t_j$; Mouschovias 1991].

During the later stages of the evolution ($ t \simgt t_1$), the core
develops a uniform central region [of radius $\simeq
\lamTcrit$], which is continually shrinking in both size and mass,
and a ``tail'' of matter and magnetic field left behind by the
central region.
Inside the ``tail'' region, near power-law behavior emerges,
with $d \ln \rhon/ d \ln r \approx -2$, $d \ln \sign / d \ln r \approx -1$,
and $ d \ln \Beq / d \ln r \approx -1$ deep inside the core, and with
$d \ln \rhon / d \ln r \approx -1.5$, $d \ln \sign / d \ln r
\approx -0.7$, and $ d \ln \Beq / d \ln r \approx -0.6$
further out, where most of the core mass is contained.
As noted by Basu (1997),
for $t \simgt t_1$, the column density and $z$ component of the magnetic
field within the inner part of the core ($r \ll R_{\rm{crit}}$) are well approximated by
the relations
\begin{equation}
\label{Basueqa}
\sign(r,t) \simeq \frac{\sigma_{\rm{n,c}}(t)}{\left(1 + \left[r/2 \lambda_{\rm{T,cr}}(t)\right]^2\right)^{1/2}}
\end{equation}
and
\begin{equation}
\label{Basueqb}
\Beq(r,t) \simeq \frac{B_{z,\rm{eq,c}}(t)}
{\left(1 + \left[r/2 \lambda_{\rm{T,cr}}(t)\right]^2\right)^{1/2}} ~~,
\end{equation}
where $\sigma_{\rm{n,c}}(t)$ and $B_{z,\rm{eq,c}}(t)$ are the 
values of $\sign$ and $\Beq$ in the uniform central region at time $t$
(see Figs. $1b$ and $1d$). PMF occurs when $\lambda_{\rm{T,cr}}(t) \rightarrow 0$.
In the inner core the magnitude of the retarding magnetic force
decreases relative to the gravitational force, becoming comparable
to that of the thermal-pressure force. Further out, at radii
$\sim \rcrit$, the magnetic force is the primary source of
support against self-gravity.

By the time $t_{10}$ the core has a radius
$R_{\rm{crit}}(t_{10})=2.59 \times 10^{-2}\rzero
= 0.111~\rm{pc}$ and a mass $M_{\rm{core}}=5.37~\msol$.
These values are in excellent agreement with typical observations
of dense ammonia cores in star-forming molecular clouds (e.g., 
Ladd, Myers, \& Goodman 1994). Separation of the magnetically supercritical
core from the massive, subcritical envelope is demonstrated in Figure
$1h$, which shows that the mass infall rate $\mdot$
is reduced by more than four orders of magnitude for $ r > \rcrit$. This
is due to the effective freezing of the magnetic field in the neutrals
at large radii, which is a consequence of the comparatively
high degree of ionization brought about by
the penetration of the external UV radiation field into the optically
thin cloud envelope (see CM95 for a discussion).
\footnote{The outer envelope may be supported even better than we
have calculated in view of the fact that
the UV photoionization rates could be higher than those listed in
Table 2, in which case the fractional ionization values in the outer
regions of the cloud (corresponding to visual extinctions $\simlt 4$) would
be slightly larger (Ruffle et al. 1998).}

We note the following results of interest for the pre-PMF phase of evolution.
First, as can be seen from Figures $1f$ and $1g$, the magnitude of the
infall speed of the neutrals in the inner regions of the collapsing core
is comparable to the isothermal speed of sound $C$
(and thus also to the fast-magnetosonic speed $\vnms = [C^2 +
\van^2]^{1/2}$, where $\van=\Beq/[4 \pi \rhon]^{1/2}$ is the
{\Alf} speed, since $\vnms$ is nearly equal to $C$ in these regions).
Moreover, at time $t_{10}$, the inward acceleration $\an$ of the
neutrals lies in the range $0.25 \gr  - 0.5 \gr$ in the inner
regions of the core,
and $|\vn|$ is a fraction $\sim 0.55$ of the free-fall speed
($\simeq 1.98~C$) at $r=\rinner$ (see \S 3.3).
Hence the core is indeed dynamically collapsing
(although not freely falling), so, as we noted in \S~1,
treating a supercritical core as being in near hydrostatic
equilibrium at PMF is {\em not} a valid approximation for
molecular cloud cores that form and evolve by ambipolar diffusion.

Second, we note the increase of the drift speed $\vd$ in the inner
flux tubes for $t \simgt t_6$ (see Fig. $1h$), which indicates
that ambipolar diffusion can continue to operate even during
the dynamical collapse of the core. This is primarily due to the
phenomenon of {\em ion depletion}, in which ions become increasingly
attached onto grains at higher densities, thereby reducing
the relative abundance of ions in the gas phase. 
This, in turn, results in a higher diffusion rate
(see CM93, CM94) and leads to an increase (for $t > t_6$) of
the mass-to-flux ratio in the flux tubes that thread the inner
core (see Fig.  $1c$). (The effect of ambipolar diffusion on the
mass-to-flux ratio during the approach to PMF is also discussed in Basu 1997.)
As we show in the next subsection (see also \S~1), ambipolar diffusion
in the interior
flux tubes is increased dramatically after PMF and
plays a significant dynamical role in the subsequent evolution of the core.

Finally, we notice that by the time $t_{10}$, $Z/r < 1$ throughout the
whole cloud, except for a very small inner region of the core
($r/\rzero \simlt 10^{-7}$; see Fig. $1i$). Note that $Z/r > 1$ does
not necessarily indicate an inconsistency in our use of the thin-disk
approximation to model clouds. As discussed in CM93 and CM95,
this approximation will be valid so long as any scalar quantity
$f(r)$ (such as $\nn$, $\sign$, $\Beq$, etc.)
satisfies the condition $f(r,t)/|\partial f/\partial r| \geq Z(r,t)$. Examination
of Figures $1a$, $1b$, and $1d$ reveals that this condition is satisfied even
in the uniform-density central region because $\partial f/\partial r \approx 0$
for $r \rightarrow 0$ before PMF.
As it turns out, the innermost region of the core, which at later
times happens to contain
densities $\nn > 10^{11}~\cc$ (see Fig. $1a$), is in any case
excluded from our calculation of the post-PMF core evolution for
the reasons outlined in \S~2.2.
\subsection{Point-Mass Formation and Protostellar Accretion Phase of
Core Evolution}
The physical data corresponding to the model at time $\tint=t_{10}$
are used as input for the protostellar-accretion phase of the
calculation. (Our results for the post-PMF epoch are relatively
insensitive to the particular
value of $\tint$ we choose; for instance, we find that the resulting
evolution differs only marginally if we use $\tint = t_8$ instead.)
The boundary of the central sink cell is taken to be $\rinner=
8.24 \times 10^{-6}\rzero =7.31~\rm{AU}$.
(The post-PMF evolution is also relatively insensitive to the
value of $\rinner$ so long as $\rinner$ is much smaller than the
core radius $R_{\rm{crit}} \approx 0.1~\rm{pc}$; for instance, the results for
$\rinner \simeq 20~\rm{AU}$ are very similar to those of our typical run.)
The density at this
position and time is $5.0 \times 10^{10}~\cc$ (see Fig. $1a$), whereas
the mass contained in the central cell at this time
is $\mcent(t_{10})= 8.5 \times 10^{-4} \msol$, which is negligible
in comparison with the total core mass $M_{\rm{core}}=5.37~\msol$.
Similarly, $\phicent(t_{10})=6.72 \times 10^{26}~\rm{Mx}$, which
is several orders of magnitude smaller than the total core
magnetic flux $\Phi_{B,\rm{core}}=1.72 \times 10^{31}~\rm{Mx}$.

The evolution of the mass in the central sink cell is shown in Figure $2a$
as a function of $\delt = t - t_{10}$.
Point-mass formation, signaled
by rapid growth of $\mcent$ with time, takes place at $\delt \leq 200~\rm{yr}$.
\footnote{Therefore, for $\delt \simgt 10^3 ~\rm{yr}$, the quantity $\delt$
can be considered to be the time elapsed since PMF. This makes our
time unit $\delt$ the same as the time $t$ typically used
in previously published self-similar collapse models (see references cited 
in \S~1), which set $t=0$ at PMF.} ---  because of the nonzero
value of $\rinner$, we can only place an upper limit on this
time. We terminate the simulation at $\delt=1.53 \times
10^5~\rm{yr}$, when $\mcent=1 \msol$. 

The accretion rate $\mcentdot$ onto the protostar (in
units of $\msol~\rm{Myr}^{-1}$) is shown in Figure $2b$ as a
function of $\delt$ ({\it solid} curve).
As in the previous (nonmagnetic and
magnetic) simulations of dynamical collapse cited in \S~1, $\mcentdot$
rapidly increases imediately after PMF due to the enhancement of the
gravitational acceleration brought about by the appearance of
the protostar.
$\mcentdot$ reaches a maximum value $\simeq 9.4~\msol~\rm{Myr}^{-1}$, which is
less than the theoretical upper limit of $13 C^3/G = 20~\msol~\rm{Myr}^{-1}$
estimated by Basu (1997) for collapsing cores in which ambipolar
diffusion remains operative.
$\mcentdot$ decreases for $\delt \simgt 4 \times 10^3~\rm{yr}$.
As we show below, the accretion rate during this time is
strongly affected by an ambipolar diffusion-induced hydromagnetic
shock that propagates outward and slows the infall of matter at larger
radii. We also plot in Figure $2b$ the central accretion rate
of a collapse model ({\it dashed} curve) that began with the same
initial data as our typical model, but with $\vd$ set equal to zero for
$\delt \geq 0$ (i.e., for this model, as discussed in \S~2.2, the
magnetic flux was artificially forced to remain frozen into the neutrals
during the post-PMF epoch). Finally, in Figure $2c$, we plot
$\mcentdot$ (in the same units as in Fig. $2b$) as a function of
$\mcent/\msol$. The {\it dashed} curve again displays the
accretion rate for the frozen-flux model.

Comparing the two models shown in Figures $2b$ and $2c$, we note
that $\mcentdot$ is the
same for $\delt<10^3~\rm{yr}$: this has to do with the fact that the
ambipolar diffusion timescale $\tad$ in the typical model is much
longer than the gravitational contraction timescale $\tgr$
during this period (see Figs. $3a$
and $3b$), so that diffusion does not greatly alter the evolution.
However, for $\delt > 10^3~\rm{yr}$, $\tad \simeq \tgr$, so
ambipolar diffusion is much more effective and leads to a redistribution
of the mass in the inner flux tubes of the core (see below). This,
in turn, {\em results in a reduction of the mass accretion rate
onto the central protostar.} Specifically, by the end of the
simulation, $\mcentdot$ in the ambipolar-diffusion model is
$\simeq 60\%$ of the accretion rate in the frozen-flux case.
The ambipolar diffusion-induced decrease in $\mcentdot$ comes on
top of the monotonic decline exhibited in the flux-frozen case
for $\delt \simgt 10^4~\rm{yr}$. The behavior of the frozen-flux model
is consistent with the results of other
nonmagnetic and magnetic collapse calculations (e.g., Hunter 1977; Foster \&
Chevalier 1993; Tomisaka 1996; Safier et al. 1997; Li 1998). As
discussed by Basu (1997, \S~5), the post-PMF decline in $\mcentdot$ is
due to the fact that, at the time of PMF, the outer mass shells of the core
are not as strongly accelerated inward (and are therefore
moving more slowly than the inner mass shells) because of their much
larger initial distance from the central point mass.
The maximum values of $\mcentdot$ in both of these numerical
models agree with those found in the self-similar magnetic collapse
solutions presented in Contopoulos, Ciolek, \& K\"{o}nigl (1997, hereafter
referred to as CCK97).

We now show that, during the post-PMF epoch, ambipolar diffusion
operates effectively in the inner core even as it continues to undergo
dynamical collapse. As we did in \S 1, we again
examine the ambipolar diffusion timescale $\tad = r/\vd$,
which we now express in a manner similar to that of Mouschovias (1989,
1991).
The force (per unit area) on the neutrals is 
\begin{equation}
\label{rforceeq}
\sign \an = \sign \gr - C^2 \frac{\partial \sign}{\partial r} +
\frac{\sign}{\tni}\vd\ ,
\end{equation}
where $\an = d \vn/dt$ is the acceleration of the neutrals.
The first term on the r.h.s. of equation (\ref{rforceeq}) is
the gravitational force (including both self-gravity of the gas and the
gravitational field of a central point mass), the second term is the
thermal-pressure force, and the last term is the frictional force due
to collisions between neutrals and ions. Dividing this equation by $r$,
one obtains
\begin{equation}
\label{tadeq}
\frac{\tad}{\tgr} = \frac{\left(\tgr/\tni \right)}
{\left[1 - \left(\tgr/\tacc\right)^2 - \left(\tgr/\ts\right)^2
\right]}\ ,
\end{equation} 
where $\tgr$ ($\equiv [r/\left|\gr\right|]^{1/2}$) is the gravitational contraction timescale
(referred to as the {\em dynamical timescale} in CM94 and CM95),
$\tacc$ ($\equiv [r/\left|\an\right|]^{1/2}$) is the acceleration
timescale, and $\ts$ ($\equiv [(C^2/r) \left|\partial \ln \sign
/ \partial  r \right|]^{-1/2}$)
is the sound crossing time. Equation (\ref{tadeq}) is the same as
equations (19) and (20) of Mouschovias (1989), except that we use the
gravitational contraction timescale $\tgr$ instead of the free-fall
time $\tff$.
Now, during the later stages of dynamical collapse, the ratio
$\tgr/\tacc$ in the denominator approaches unity and tends to increase
$\tad/\tgr$ (the ratio $[\tgr/\ts]^2$ is negligible).
However, effective diffusion (corresponding to $\tad/\tgr \simlt 1$) can
still occur if the ratio $\tgr/\tni$ in the numerator of equation
(\ref{tadeq}) is small during this period --- due either
to a decrease in $\tgr$ or to an increase in $\tni$. Hence, if
$\tgr/\tni \ll 1$, ambipolar diffusion can still redistribute mass and
flux in a dynamically collapsing core.
\footnote{The criterion $\tgr/\tni \ll 1$ is equivalent to 
requiring $\lambda_{\rm{M,cr}}/\lambda_{\rm{A}} \ll 1$, where $\lambda_{\rm{M,cr}} \approx \van \tff \approx \van \tgr$
is the critical magnetic lengthscale, and $\lambda_{\rm{A}} \approx \van \tni$
is the {\Alf} lengthscale, as defined by Mouschovias (1991). The
condition $\lambda_{\rm{M,cr}} \ll \lambda_{\rm A}$ for effective
ambipolar diffusion during rapid core collapse was put forth by Mouschovias (ib., \S~4.1).}
That this is indeed the case in our collapse model can be seen in Figures $3a$ and $3b$, which
show, as functions of $\delt$, $\tad/\tgr$, $\tgr/\tni$, and the total
(nondimensional) magnetic flux contained within cells 1 and 5 of our
computational mesh. Examination of these figures reveals that
$\tgr/\tni$ decreases with increasing $\delt$. This is caused solely by
the decrease in the value of $\tgr \propto |\gr|^{-1/2} \propto \mcent^{-1/2}$
(see eq. [\ref{newgreq}]). On the other hand, $\tni \propto \nni^{-1}$
(see eq. [\ref{tnieq}]) changes very little during this period
due to the fact that, as noted in \S~3.2, $\nni \simeq const$ for $\nn
\gg 10^7~\cc$. The rapid increase of $\mcent$ with $\delt$ following
PMF (see Fig. $2a$) thus leads to a strong decrease in $\tgr/\tni$,
resulting in a ``revitalization'' of ambipolar diffusion in the inner
flux tubes during this phase. For $\delt \simgt 10^3~\rm{yr}$,
$\tad/\tgr \simeq 1$, and {\em the growth of the magnetic flux contained
within these two cells is halted}. The flux then begins to ``pile up''
in the inner cells of the computational mesh. 
(The revitalization of ambipolar diffusion after PMF can also be
understood in terms of the scaling of the magnetic field following point
mass formation; see Appendix C.)

The effect that the piled-up magnetic flux has on the inner flux tubes
of the typical model can be seen in Figures $4a$ and $4b$, which
display, respectively, $\Beq/\Beqcb$ and $\sign/\sigcb$ as functions of
$\delt$ for cells 2, 5, 10, 15, 20, 25, and 27 of the computational
mesh. Figure $4a$ shows that the piling up of magnetic flux begins 
close to the axis of symmetry and then, with increasing time,
becomes noticeable at progressively larger radii. Hence, after PMF,
{\em ambipolar diffusion
causes a front of magnetic flux to move outward from the innermost flux
tubes}. Ambipolar diffusion is so effective behind this {\em hydromagnetic
disturbance} (HMD) that the inward advection of magnetic flux stops
--- with the field lines being held in place or else moving slowly
outward, in either case with $|\vi| \ll |\vn|$.
This behavior is reproduced by the self-similar collapse solutions
of CCK97, which also incorporate the effect of ambipolar
diffusion on the post-PMF core evolution. The magnetic
field strength $\Beq$ is greatly enhanced after the passage of the HMD
in each cell.
The effect of the HMD on the behavior of the column density
varies with the distance from the center. In particular, the
column density in cells 2 and 5 shows no effect from the passage of the
magnetic field front (see Figs. $4a$ and $4b$).
This is because ambipolar diffusion is so rapid in these cells that
very little of the increased magnetic force from the enhanced field
is transmitted to the neutrals
through ion--neutral collisions. However, after the HMD reaches
larger distances, the collisional coupling
between the two fluids increases (see Fig. $6h$). Furthermore,
the piling-up of flux increases the local field strength
at these distances by a larger relative factor.
As a result of these two effects, the
increased magnetic force acting on the neutrals is able to temporarily
{\em decelerate} the infalling matter (from slightly supersonic to
subsonic infall speeds; see Fig. $6e$) for cells with index $l \geq 8$.
For these cells, $\sign$ {\em increases} immediately
after the passage of the front. The effect of the HMD at
larger radii can be described in terms of a {\em hydromagnetic
shock that propagates in the weakly ionized gas}. (As noted in \S~1,
the formation of such a shock was predicted by Li \& McKee 1996; we
further analyze the nature of this C-type shock in \S~4.2.)
After traversing the shock, the neutrals are gradually
reaccelerated and eventually reach free-fall speeds.
>From Figure $4b$ we find that $\sdelt \equiv d \ln \sign/d\ln \delt$
lies in the range
$-0.76 \simlt \sdelt \simlt -0.55$, which can be compared with
the value $\sdelt = -1/2$ that characterizes self-similar collapse models of
thin, axisymmetric disks (Li \& Shu 1997; CCK97).

The position of the hydromagnetic disturbance $\rfront(t)$
in units of the cloud radius $\rzero$ is presented in Figure $5a$ as a
function of $\delt$. The speed $\vfront$ ($=d \rfront/dt$, in
units of $C$) of the front in the reference frame of the central
protostar is shown in
Figure $5b$. (The time intervals between successive data points
in these figures are larger than those in Figures 2--4. Figures
$5a$ and $5b$ therefore show a time-averaged motion of the HMD
and do not exhibit the steady, shock-induced oscillations seen
in Figures 2--4 at later times.)
Early on, the radially outward motion of the front is
nonsteady, and depends on the rate at which the flux external to the
front enters and piles up in the region of rapid ambipolar diffusion.
At later times, $\vfront < C$. The Mach number of the hydromagnetic
disturbance in the reference frame of the upstream gas,
${\cal M}_{\rm{HMD}} = (\vfront - \vnu)/C$ (where $\vnu$ is the
infall speed of the neutrals just ahead of the front), is shown
in Figure $5c$ as a function of $\delt$. The motion of the HMD
relative to the upstream gas is supersonic, and, for $\delt
\simgt 10^4~\rm{yr}$, ${\cal M}_{\rm{HMD}} \simeq 2.8$.
It is also of interest to note that, in the vicinity of the front, the
fast-magnetosonic speed $\vnms$ is typically $\simlt 2 C$. Hence the
hydromagnetic Mach number ${\cal M}_{\rm{ms,HMD}}=(\vfront-\vn)/\vnms$
of the HMD relative to the upstream neutrals is also $> 1$.

Spatial profiles of various quantities in the typical model are
shown in Figures $6a$--$6i$ as functions of $r/\rzero$, at seven
different times $\delt$. As noted earlier, $\delt$ is essentially
the time elapsed since PMF. For this typical model, $\delt_{0}=0$,
$\delt_{1}=2.17 \times 10^2~\rm{yr}$,
$\delt_{2}=6.11 \times 10^2~\rm{yr}$, $\delt_{3}= 1.54 \times 10^3~\rm{yr}$,
$\delt_{4}=3.81 \times 10^3~\rm{yr}$, $\delt_{5}=2.38 \times 10^4~\rm{yr}$,
and $\delt_{6}=1.48 \times 10^5~\rm{yr}$. An {\em asterisk} on a curve
locates, as in Figures $1a$--$1i$, the instantaneous position of the
critical flux tube $\rcritb$. The core radius does not change
from its initial value of $R_{\rm{crit}}(\delt=0)=R_{\rm{crit}}(t_{10})
=2.59 \times 10^{-2} \rzero = 0.111~\rm{pc}= 2.30 \times 10^4~\rm{AU}$,
providing further evidence for effective core--envelope separation, 
as discussed in earlier models presented by Mouschovias and coworkers
(see references in \S~1).

Figure $6a$ shows the total mass $M$ in
$\msol$. For $\delt_j > 0$ the curves flatten for small radii, revealing
point-mass formation since $M(r)/r \neq 0$ for $r \rightarrow 0$.
For $r > \rcritb$, $M(r)$ changes very little as the magnetically
subcritical envelope continues to be supported by magnetic forces.
There is a small transfer of mass from the envelope to the
supercritical core during the post-PMF epoch; by the time $\delt_6$ the core
mass has increased by $8\%$ to $5.80~\msol$.
Figure $6b$ displays $\Beq$ normalized to $\Beqcb$, Figure
$6c$ shows the ratio $\sign/\sigcb$, and Figure $6d$ depicts
$\BrZ$, the $r$ component of the magnetic field at the upper surface of the
disk, normalized to $\Beqcb$. The infall speed $\vn$ of the neutrals
(normalized to $C$) is shown in Figure $6e$, the ion--neutral drift
speed $\vd$ (also normalized to $C$) is displayed in Figure $6f$, and
the mass accretion rate $\mdot$ in $\msol~\rm{Myr}^{-1}$
($=0.65 C^3/G$ for the temperature and gas composition assumed in this
model) is exhibited in Figure $6g$. The ratio $\tni/\tgr$ is
plotted in Figure $6h$, and the ratio $Z/r$ is shown in Figure $6i$.

Taken together, Figures $6b$--$6i$ show the effect of ambipolar diffusion
and the outward progression of the HMD (and the resulting hydromagnetic
shock) on the evolution of the core. For $\delt < \delt_2$, ambipolar
diffusion has not yet been revitalized, and mass and magnetic flux are
continually advected inward. During this period, $\sign$ and $\Beq$
in the core decrease with increasing $\delt$. $\BrZ$ increases
as field lines are bent inward while being carried along with the
neutrals. For $\delt_j > \delt_2$, ambipolar diffusion has become rapid
enough to halt the advection in the innermost cells. Subsequently the
front of piled-up magnetic flux expands outward to larger radii, dramatically
enhancing the value of $\Beq$ in its wake.
Bending of the interior field lines is slowed and temporarily arrested
in the region of rapid ambipolar diffusion behind the HMD for
$\delt_2 \simlt \delt \simlt \delt_5$ (the behavior for $\delt > \delt_5$
is considered below). Further out from the center, the
magnitude of the local gravitational field strength decreases while the 
degree of ionization increases, resulting in less efficient ambipolar
diffusion and more effective collisional coupling between the ion and
neutral fluids.
At times $\delt \simgt \delt_4$ the HMD is propagating
out into the regions of the core ($r/\rzero \simgt 10^{-4}$) where
$\tni/\tgr \simlt 1$ (see Fig. $6h$).
At these radii both the
collisional coupling of the ion and neutral fluids and the
enhanced magnetic field strength behind the HMD are
large enough to affect the inflow of the neutrals, and a shock
forms. Our numerical calculation thus confirms the evolutionary
sequence originally outlined in Li \& McKee (1996).
A local maximum in $\sign$ occurs because the neutrals are ``hung up''
(i.e., decelerated; see Fig. $6e$) in the region of stronger magnetic
field strength, and their infall is {\em interrupted} as they are forced to
diffuse through the stationary, or slowly expanding, ions and field
lines contained in the HMD ($\vd \approx |\vn|$ and $|\vi| \ll |\vn|$
behind the HMD; see Fig.  $6f$). After the neutrals diffuse through the
shock, they resume dynamical collapse toward the origin: at time
$\delt_6$, $|\an| \simgt 0.90 |\gr|$ for $r \simlt 4 \times 10^{-5} \rzero
= 35.5 ~\rm{AU}$, $\sign \propto r^{-1/2}$, and $\vn \propto r^{-1/2}$
(which is the characteristic behavior of free-fall collapse induced 
by a central point mass, also found in self-similar models of
gravitational collapse in isothermal disks; Li \& Shu 1997; CCK97).
At that time $|\vn|$ is a fraction $\sim 0.94$ of the free-fall
speed ($\simeq 76.46~C$) at $r=\rinner$.
Depletion of matter, caused by the interruption in the infall of the
neutrals at the shock front, is the reason for the local minimum in
$\sign$ that occurs between the shock front and the free-fall region.
This causes the accretion rate behind the shock to decrease for
$\delt \simgt \delt_4$.  Hence ambipolar diffusion, through the action
of the hydromagnetic shock, has {\em increased} the magnetic
field strength in the collapsing core for $r \simgt 10^{-4} \rzero$
($\simeq 88.7~\rm{AU}$) and {\em reduced}
the accretion rate onto the central protostar --- exactly the
{\em opposite} of what one would have expected a priori.

For comparison we also plot in Figures $6b$, $6c$, and $6d$
({\it dashed} curves), for the model in which flux freezing into
the neutrals was assumed to hold throughout the post-PMF epoch,
the profiles of $\sign$, $\Beq$, and $\BrZ$, respectively, at
the time $\delt_6$ the mass of the central protostar reached 
$\mcent = 1 \msol$. (For this
model, $\delt_6 = 1.23 \times 10^5~\rm{yr}$.)
In the absence of ambipolar diffusion, the magnetic flux in the core is
continually advected into the central sink, and $\Beq$ is greatly
reduced from the value that characterizes the core of the typical
ambipolar-diffusion model (see Fig. $6b$); $\BrZ$, on the other hand, is
much larger in the core of the frozen-flux model (see Fig. $6c$)
because the field lines continue to be bent inward as mass and flux are
advected into the central sink. Examination of Figure $6c$ reveals that
the column density of the ambipolar-diffusion model is significantly
larger than that of the frozen-flux model for
$r \simgt 1.6 \times 10^{-4}~\rzero=142~\rm{AU}$,
i.e., the region just behind the shock front. For $r \simlt 142~\rm{AU}$
the opposite occurs: $\sign$ in the model that
includes ambipolar diffusion is much smaller than in the frozen-flux case. 
This is
because more of the core mass remains further away from the center
in the ambipolar-diffusion model, being {\em trapped} at larger
radii by the pressure of the enhanced magnetic field in the wake of
the HMD. {\em Hence, ambipolar diffusion, by giving rise to a hydromagnetic
shock, has significantly altered the evolution of the column density
profile in the core.} Moreover, a hydromagnetic shock
does not occur in the core of the frozen-flux model. This is contrary
to the results of the self-similar model of Li \& Shu (1997). This
discrepancy arises because the initial conditions assumed by Li \& Shu
correspond to a stationary configuration rather than to the
dynamical one indicated by the numerical simulations
of the pre-PMF phase of collapse (see \S~3.2). In fact, as discussed
in CCK97, shocks also do not form in the absence of ambipolar diffusion
in self-similar models that do not use the initial
singular-isothermal-disk setup of Li \& Shu.
Finally, we note that, for $r \simgt R_{\rm{crit}}(\delt)$, the
magnetic field is the same in both models, reflecting the fact that
ambipolar diffusion is so ineffective in the massive, subcritical
envelope that the field there can be considered to be frozen into the
neutrals at all times.

Figure $6b$ shows that, by the time $\delt_6$, $\Beq$ has been
significantly reduced for $r \simlt 2.5 \times 10^{-4} \rzero = 222~\rm{AU}$.
This is due to the fact that the drift speed $\vd(r)$ exceeded the
critical value $\vdcrit \approx 12 ~\rm{km}~{\rm s}^{-1}$ in this
region of the core for $\delt_5 < \delt < \delt_6$. Therefore, during
this period, the neutral--ion collisional force scaled quadratically with
$\vd$ rather than linearly (compare eqs. [\ref{newdrifteq}] and
[\ref{rforceeq}]), so the frictional force between the ions and the
neutrals was significantly increased in this part of the core.
This had the effect of {\em refreezing} the magnetic flux into
the neutrals in the innermost computational cells, resulting in the field
being dragged by the collapsing neutrals into the central sink
and therefore decreasing in strength in these cells.
However, further out ($r > 2.5 \times 10^{-4}~\rzero$), where
$\vd(r) < \vdcrit$, the
bulk of the flux front remains largely unaffected by the collapse of its
``floor.'' By $\delt_6$, though, $\vd$ has fallen below $\vdcrit$ (see
Fig. $6f$) even in the innermost cells of the computational mesh.
Bending of the field lines resumes in this region of the core at that
time, with a subsequent increase in $\BrZ$ (see Fig. $6d$).

We note that beyond the core radius [i.e., for $r \geq R_{\rm{crit}}(\delt) \simeq 2.52 \times 10^{-2}~\rzero=0.11~\rm{pc}$]
$|\vn| \ll C$ (see Fig. $6e$),
reflecting effective support of the massive, subcritical
envelope by magnetic forces. Ambipolar diffusion in the envelope is so
ineffective (resulting from the much greater relative
abundance of ions, due to ionization by the external UV radiation
field) that $\vd \ll C$ (see Fig. $6f$), and the magnetic field there
can be considered
to be frozen into the neutrals; furthermore, $\BrZ \ll \Beq$ in the
envelope (see Figs. $6b$ and $6d$), indicating that the field lines
there remain essentially straight and parallel. Effective
separation of the supercritical core and the envelope continues for
$\delt \geq 0$, as can be seen by the dramatic drop (by more than four
orders of magnitude) in $\mdot$ for $r \simgt \rcritb$
(see Fig. $6g$). As noted in \S~1, Safier et al. (1997)
modeled the evolution of a cloud's envelope during the post-PMF epoch.
Equation (104) of Safier et al. gives their predicted accretion rate
as a function of the initial cloud mass, radius, and
the ratio of the initial mean and central density. Evaluating their
expression for the parameters used in our typical model
($M_{\rm d} = 98.3~\msol$, $\rzero = 4.29~\rm{pc}$, and
$\langle \nn \rangle_0/\nncb=0.144$) yields a spatially uniform envelope
accretion rate $\mdot =3.7~\msol~\rm{Myr}^{-1}$, which is close to
our calculated accretion rate at the core boundary (see
Fig. $6g$). However, for $r \simgt 0.26 \rzero \simeq 0.11~\rm{pc}$
their predicted value of $\mdot$ is several orders of
magnitude higher than the one we have obtained in the envelope of
our typical model. The difference can be attributed to the fact that
Safier et al. chose to study model clouds for
which ionization by the external interstellar UV radiation field 
was unimportant, resulting in a lower degree of ionization and a
correspondingly higher ambipolar diffusion rate in the cloud envelope.

Finally, the effect that the tidal gravitational field of the central
protostar (see eq. [\ref{newforcbaleq}]) has on the vertical structure
of the cloud can be seen in Figure $6i$. For times
$\delt > 0$ the inner region of the core near the central mass
is significantly compressed by
the tidal force, and $Z/r$ becomes $\ll 1$. Further compression
will be provided by the squeezing effect of the radial magnetic field
component (see eq. [\ref{newforcbaleq}]), particularly at late
times in the vicinity of $\rinner$. All in all, the disk remains
geometrically thin at all radii $r > \rinner$ for the entire
duration of the simulation.
\section{Discussion}
\subsection{Observational Comparisons and Predictions}
We may compare our result for the protostellar accretion rate
during the PMF epoch with observations of star-forming
cores. As shown in \S~3.3, the accretion rate rises rapidly early
on to $\mcentdot \simeq 9.4~\msol~\rm{Myr}^{-1}$ for $\delt \simlt 10^3~\rm{yr}$
(see Fig. $2b$) and stays at this value up to the formation of the
hydromagnetic shock. For $\delt \simgt 4 \times 10^3~\rm{yr}$ the shock
is able to decelerate the infalling matter, and the accretion rate
decreases to $\mcentdot \simeq 5.6~\msol~\rm{Myr}^{-1}$ by 
$\delt \simeq 1.5 \times 10^5~\rm{yr}$ (the time when
$\mcent = 1~\msol$). Therefore $\mcentdot$ decreases with
increasing central mass (see Fig. $2c$).
This is consistent with estimates of ages and accretion rates
($\propto t_{\rm{age}}^{-1}$) for young stellar objects, as deduced
from evolutionary diagrams inferred from observations of Class 0 and
Class I objects (e.g., Saraceno et al. 1996).
In particular, the lifetimes of Class 0 objects were estimated in this way to
be an order of magnitude shorter than those of Class I objects,
providing evidence for a decrease in the protostellar accretion
rate as an object evolves from a Class
0 source to a Class I source (e.g., Andr\'{e} 1995; Ward-Thompson 1996).
Another argument for a time-dependent accretion rate was given by
Bontemps et al. (1996), who analyzed the observed CO momentum
flux of several young stellar objects and found a noticeable
decline in the CO flux with decreasing circumstellar envelope
mass. They suggested that this
is indicative of a decrease in the protostellar accretion rate
(which they assumed to be proportional to the mass outflow rate) as an
object evolves from Class 0 to Class I.

It is also of interest to note the ${}^{13}{\rm{CO}}(J=1-0)$
observations of infalling disks for the protostellar
candidates HL Tauri (Hayashi, Ohashi, \& Miyama 1993) and L1551-IRS5
(Ohashi et al. 1996). HL Tauri has a mass $\sim 0.6~\msol$ and
a surrounding disk with radius $\sim 1400~\rm{AU}$ and
mass $\sim 0.03~\msol$. From the observed kinematics Hayashi et al.
derive an accretion rate $\sim 9~\msol~\rm{Myr}^{-1}$ at
$r \sim 700~\rm{AU}$.
The embedded protostar L1551-IRS5 has a mass
$\sim 0.5~\msol$ and is surrounded by a disk with radius
$\sim 700~\rm{AU}$ and mass in the range $3.9 \times 10^{-2} - 8.1 \times 10^{-2}~\msol$.
Ohashi et al. deduce an accretion rate in the range
$13 - 26~\msol~{\rm{Myr}}^{-1}$ at
$r \sim 600~\rm{AU}$. These values are comparable to our model
results for $\delt_5 \simlt \delt \simlt \delt_6$
(corresponding to $0.2~\msol \simlt \mcent \simlt 1~\msol$;
see Fig. $6a$). During this period, $ 6~ \msol~{\rm{Myr}}^{-1} \simlt \mdot \simlt 9~ \msol~{\rm{Myr}}^{-1}$
for $r \simgt 500~\rm{AU}$ (see Fig. $6g$).
[Note, however, that the
temperatures of HL Tauri and L1551-IRS5 are in the range
$15 - 50~{\rm K}$, which is greater than our assumed value of
10 K and should lead to higher values of $\mdot$ and $\mcent(t)$;
e.g., Shu et al. 1987. For a discussion of how quantities scale with
temperature in our models, see Basu \& Mouschovias 1994.]
In Figure 7 we show the mass ($M - \mcent$) of the gas surrounding the
central point mass in our typical model as a function
of $r/\rzero$ for the same seven times $\delt_j$ as in Figure 6.
(Taken together, Figs. $6a$ and 7 may be taken to represent the
evolution of a protostar
from a Class 0 to a Class I object.)
For times $\simgt \delt_5$, the surrounding disk mass spans the
range $0.01 - 0.1~\msol$ for $r \simgt 500~\rm{AU}$, which agrees
with the ${}^{13}{\rm{CO}}$ disk masses of HL Tauri and L1551-IRS5
cited above. Finally, we note that the age of the oldest part of the
molecular outflow from L1551-IRS5 is estimated to be $\sim 10^5~\rm{yr}$
(Bachiller, Tafalla, \& Cernicharo 1994). This age is consistent with
the time $\delt$ needed for the central mass in our typical model to
become $\simgt 0.3~\msol$ (see Fig. $2e$).

Another observational consequence of our model is the
magnetic field structure in the core after PMF. Figures $6b$ and $6d$
show that $\BrZ \approx \Beq$ inside the core for the radius range
$ 2 \times 10^{-4} \simlt r/\rzero \simlt 4 \times 10^{-3}$.
Hence, there is significant curvature of field lines
(though, as discussed in \S~3.3, there is less bending than there
would be if the field had remained frozen into the neutrals), with
bending angles $\theta_B \approx \arctan(\BrZ/\Beq)$ in the
range $20^{\circ}- 50^{\circ}$ for $180~ {\rm AU} \simlt r
\simlt 3.5 \times 10^3~{\rm{AU}}$. 
\footnote{In agreement with the results of CM94 and CM95, we find that
the magnetic tension force is generally not negligible in comparison
with the magnetic pressure force at any radius (in either the core or the
envelope), both before and after PMF.}
This type of field geometry may be described roughly as having an
hourglass shape. In fact, sub-mm polarimetry of the cores
of W3 IRS5 (Greaves, Murray, \& Holland 1994), Mon R2 (Greaves, Holland,
\& Murray 1995), and OMC-1 (Schleuning 1998) find field geometries
suggestive of an hourglass shape on sub-parsec scales. In contrast,
field lines remain essentially straight and parallel in the
magnetically supported envelope ($r > 0.1~\rm{pc}$), even after PMF.
This agrees with polarimetric observations of molecular clouds that
indicate well-ordered fields on these scales (e.g., Hildebrand,
Dragovan, \& Novak 1984; Hildebrand 1989, 1996; Novak et al. 1989;
Kane et al. 1993; Hildebrand et al. 1995).

A unique prediction of our model is the large ion--neutral
drift speed that occurs during the post-PMF epoch. As shown in Figure $6f$,
effective ambipolar diffusion following PMF yields $\vd \approx |\vn| \gg C$
in the inner regions of the core. In our typical model we find
$\vd \simgt 1~\rm{km}~{\rm s}^{-1}$
for $\delt \simgt 2 \times 10^4~\rm{yr}$ on scales
$r \simlt 2 \times 10^{-4} \rzero \simeq 180~\rm{AU}$.
Large drift speeds between neutrals and ions (such as $\HCOp$,
$\rm{HCN}^+$, $\rm{DCO}^+$, to name but a few) on these scales are
therefore expected in our model. Detection of such drift speeds 
(through high-resolution observations of HCN or $\HCOp$, say) could
be used to observationally confirm our model results and to
distinguish them from those of nonmagnetic collapse calculations
(e.g., Shu 1977, Hunter 1977, Foster \& Chevalier 1993) or of
magnetic collapse models that do not account for the effect of
ambipolar diffusion (e.g., Tomisaka 1996; Li \& Shu 1997).
\subsection{Features of the Hydromagnetic Shock}
The properties of hydromagnetic
shocks in partially ionized gases have been developed extensively
by many other authors (e.g., Mullan 1971; Draine 1980; Chernoff 1987;
Roberge \& Draine 1990; Draine \& McKee 1993; Smith \& Mac Low 1997).
Because the ion {\Alf} speed $v_{\rm{A,i}} = \Beq/(4 \pi \mi \nni)^{1/2} = (\mn/\mi)^{1/2} (\nn/\nni)^{1/2} \van$
is much larger than $\van$, $|\vn|$, and $|\vi|$ in our model, we
expect the outward-propagating shock that develops after PMF to have 
a magnetic precursor. This is indeed what we find in our simulation:
the jump in the ion speed $\vi$ and the magnetic field strength $\Beq$
typically occurs at a distance of 1 to 3 computational mesh spacings
further away from the symmetry axis than the jump in the neutral infall
speed $\vn$ and the column density $\sign$. The displacement between
the locations of the head of the disturbance in the neutral fluid and
in the plasma and magnetic field decreases at later times. Examination
of Figures $5b$, $5c$, and $6e$ reveals that, in the reference frame of
the shock, the preshock infall speeds are supersonic, and the
postshock speeds are also supersonic or just slightly subsonic.
Similarly, the ion infall speeds (in the frame of the shock) are
much less than the ion {\Alf} speed. Therefore the shock we observe in
our model is probably best classified as being C-type.
\footnote{A C-type shock is characterized by neutral velocities
that (in the shock frame) remain supersonic
throughout. Hence the shock in our simulation cannot be
strictly of this type when the downstream neutral speed is
subsonic. In that case a viscous (J-type) subshock may form
(Draine \& McKee 1993), although, as noted by Li \& McKee
(1996), turbulent diffusivity behind a real shock could plausibly
keep the postshock flow supersonic and thereby obviate the need
for such a subshock.}
Making the approximation that in the vicinity of the
shock the predominant magnetic stress is that due to the magnetic
pressure gradient, the ion force equation becomes
\begin{equation}
\label{ionforceq}
\frac{\sign}{\tni} \vd = - \frac{Z}{4 \pi} \frac{\partial \Beq^2}{\partial r}
\end{equation}
(see eqs. [28c] and [51] in CM93, which contain additional
terms, involving in particular the magnetic tension force, that
could be used to refine the following simple estimate). This yields an
approximate shock width 
\leteq
\begin{eqnarray}
\label{shkapproxa}
\delshk &\approx& 
\frac{\Bequ^2}{4 \pi \sign} \frac{\tni Z}{\vd} \left[\left(\frac{\Beqd}{\Bequ}\right)^2 -1 \right] = C \tni \frac{\left(\vanu/C \right)^2}{2 \left(\vd/C\right)} \left[\left(\frac{\Beqd}{\Bequ}\right)^2 -1 \right] \\
\label{shkapproxb}
&=& 7.8 \times 10^{13} \frac{\left(T/10~{\rm K}\right)^{1/2}\left(\vanu/C\right)^2}{\left(\mn/2.33~{\rm{a.m.u.}}\right)^{1/2}\left(\nni/0.1~\cc\right)\left(\vd/C\right)}
\left[1 + 0.067 \frac{\left(\mHII/2~{\rm{a.m.u.}}\right)}{\left(\mi/30~{\rm{a.m.u.}}\right)}\right] \nonumber \\
&&\hspace{5em}\times
\left[\left(\frac{\Beqd}{\Bequ}\right)^2-1\right]~\rm{cm}\ ,
\end{eqnarray}
\beq
where $\Bequ$ and $\Beqd$ are the values of $\Beq$ upstream and
downstream of the shock, and $\vanu$ is the upstream {\Alf} speed.
In deriving the last equality of equation (\ref{shkapproxa}) we have
used the relation $\sign= 2 \rhon Z$; equation (\ref{tnieq}) and the
relation $C= \left(\kB T/\mn\right)^{1/2}$ have been used in deriving
equation (\ref{shkapproxb}). At the time $\delt_6$ the shock front is
located at
$r \simeq 3.9 \times 10^{-3} \rzero = 5.2 \times 10^{16}~{\rm cm}$,
and, in the vicinity of the front, $\nni \simeq 10^{-2}~\cc$, $\vanu \simeq C$,
$\Beqd/\Bequ \simeq 4.6$ (see Fig. $6b$), and $\vd \simeq 0.4 C$ (see Fig. $6f$).
For these values our rough estimate for the shock width given by
equation (\ref{shkapproxb}) yields $\delshk \simeq 4.3 \times 10^{16}~{\rm{cm}}$.
Examination of Figure $6e$ at the time $\delt_6$ reveals that the shock has
an actual width $\delshk \simeq 2.1 \times 10^{-3} \rzero = 2.8
\times 10^{16}~\rm{cm}$.
(Thus $\delshk/r_{\rm shk} \approx 0.5$
at that time, so the thin-shock approximation that underlies
the estimate [17] is marginally satisfied.)

As noted in \S~1, Li \& McKee (1996) proposed that a hydromagnetic
shock would form in a collapsing core as a result of the decoupling of
flux from the ion and neutral fluids because of Ohmic dissipation (a
process that becomes important in regions of density $\nn \gg 10^{11}~\cc$;
see \S~2.1). They argued that the accumulating flux diffuses outward to
regions of lower density, where improved coupling with the matter
causes it to present an obstacle to the infalling neutral gas --- thereby
giving rise to a hydromagnetic shock. While our numerical
results have confirmed Li \& McKee's basic shock-formation scenario, they have
revealed that ambipolar diffusion, which mediates the shock,
can also, following PMF, halt the inward advection of magnetic
flux on scales ($r \simgt 5~\rm{AU}$) where Ohmic
dissipation is not important. In other words, our results have
shown that, during the
post-PMF epoch, the field--matter decoupling that drives the
hydromagnetic shock is due to ambipolar diffusion alone and does
not depend on the effect of Ohmic dissipation at $r < \rinner$.

Because of the similarity in the basic
shock-formation mechanism, it is of interest to compare our detailed
simulation results with the predictions of the (simplified and analytic)
shock model of Li \& McKee (1996). From their requirement that
the magnetic pressure of the shock balance the ram pressure of
the neutrals, which were assumed to be freely-falling into the shock, Li
\& McKee derived relations for the shocked magnetic field
strength and the shock location (see their eqs. [7] and [8]) in terms of the
accretion rate $\mdot$, the flux-to-mass ratio
in units of the critical value for collapse (dubbed $\epsilon$ in
their paper), a parameter related to the logarithmic gradient of the
magnetic field (dubbed $\chi$), the ratio $Z/r$ of the local gas scale
height and the radius (dubbed $h$), and the protostellar mass (denoted by
$m_{\ast}$).  At the time $\delt_6$ we have at the location of our shock
$\dot{M} \approx 9~\msol~{\rm{Myr}}^{-1}$,
$\epsilon \approx 0.9$, $\chi \approx 2$,
$h \approx 0.3$, and
$m_{\ast}=\mcent(\delt_6) \approx 1~\msol$. Inserting these values into
their equations (7) and (8) yields
a shocked magnetic field strength $\sim 630~\mu{\rm G}$ and a shock radius
$\sim 2.1 \times 10^3 ~{\rm{AU}}$; by comparison, in our model
the shocked magnetic field
strength at that time is $\Beqd \approx 330~\mu\rm{G}$ and the shock
radius is $r_{\rm shk} \approx 3.5 \times 10^3~\rm{AU}$
(see Figs. $6b$ and $6e$).
The main reason why the analytic expression overestimates the
numerically calculated magnetic field strength
is that, contrary to the assumption of Li \& McKee,
the preshock neutrals are not in free fall but, rather, are
strongly decelerated by magnetic forces (in our simulation we find that
the preshock acceleration of the neutrals is reduced to $\simeq
0.25 \gr$). The corresponding reduction in the preshock ram
pressure leads to a lower value of the postshock field
amplitude, with a further reduction in the calculated field
strength brought about by the contribution of magnetic tension 
(ignored in the analytic estimate) to the total magnetic
force. Since the analytic estimate of $r_{\rm shk}$ is based on
relating the postshock field strength to the total magnetic flux
inside the shock, the overestimate of the field strength
naturally results in an underestimate of the shock radius.
Despite these discrepancies, Li \& McKee's analytic representation
of the shock parameters provides a decent
approximation to the results of our numerical calculation.

Comparison of the column density $\sign$ (see Fig. $6c$) and
neutral infall speed $\vn$ (see Fig. $6e$) behind the shock shows
qualitative agreement with Figures $2a$ and $2d$ of Li \& McKee (1996),
including the free-fall behavior near the central protostar. In the
pre-shock region, however, our infall speeds are smaller than theirs
because of their assumption of free fall upstream of the shock:
Li \& McKee typically overestimate $\vnu$ by a factor $\sim 2$.
As a result, the Mach number of the shock relative to the
upstream gas (and thus the shock strength) is greater in their model
than in ours (see Fig. $5c$) by a similar factor. We have not compared
our results for the magnetic field structure behind the shock with those
of Li \& McKee on account of the fact that their system of MHD equations
was not closed (it did not include the magnetic induction equation),
so that they were unable to calculate the magnetic field with any
accuracy (see their Fig. $2b$).
\subsection{Stability of the Core Against Magnetic Interchange}
Our simulation has revealed that rapid ambipolar diffusion
occurs behind the outward-propagating HMD. The effect that this has on
the mass in the flux tubes downstream of the HMD can be seen
in Figure $8a$, which shows the local mass-to-flux ratio
$d M/d \phiB = \sign/\Beq$
(normalized to the critical value for collapse) as a function of
$r/\rzero$ for the same seven times $\delt_j$ as in Figure 6.
For $\delt > \delt_1$ a local minimum in $\sign/\Beq$
appears after the passage of the HMD. Hence, there
is a region behind the HMD for which $d (\sign/\Beq)/dr > 0$. This
is a necessary condition for the onset of an interchange
instability (e.g., Spruit \& Taam 1990; Lubow \& Spruit 1995; Spruit,
Stehle, \& Papaloizou 1995). Li \& McKee (1996) speculated that such a
situation could arise in the wake of a hydromagnetic shock in a
collapsing core and suggested that it would act as source of
turbulence in the postshock region of the inflow.
\footnote{Li \& McKee (1996) also noted that the shock may be
unstable to the Wardle instability, which involves ions
collecting in magnetic field ``valleys'' (Wardle 1990). However,
the shock will be immune to this instability if the ion density is
determined by the local chemical reaction balance (as assumed in
our calculation) rather than by the divergence of the ion mass flux.}

Blaes \& Balbus (1994) found that the magnetic shearing
instability in differentially rotating disks could
be stabilized if the disk is weakly ionized. This will also be the
case for the interchange instability in a weakly ionized disk: instability
is possible only if the growth rate $\gaminst$ and the neutral--ion
collision time $\tni$ satisfy the condition $\gaminst \tni < 1$. This
condition reflects the fact that there has to be sufficient collisional
coupling between the ion and neutral fluids for a magnetic interchange
instability to grow in the neutrals; otherwise the instability is
damped. Calculation of $\gaminst$ in our model
is hampered by the fact that previous studies of interchange instability
have been carried out only for disks that are in hydrostatic equilibrium,
with exact balance between magnetic and gravitational forces.
We can apply the results of these studies to our model only
if the region behind the shock where $d (\sign/\Beq)/dr > 0$ is
effectively in quasi equilibrium, with approximate balance between
gravitational and magnetic forces, and with infall speeds
$|\vn| \ll (r |\gr|)^{1/2}$ ($\simeq$ the free-fall speed).
In our model, the magnitude of the
acceleration of the neutrals in this region of the core does not become
$\simlt 0.1 | \gr |$ until times $\sim \delt_6$; hence,
approximate equilibrium between magnetic and gravitational forces
is valid only at these later times.
Spruit \& Taam (1990) found that the growth rate for the most unstable
linear interchange modes is
\begin{equation}
\label{growtheq}
\gaminst = \left( \frac{\Beq \BrZ}{2 \pi \sign} \frac{d}{dr} \ln \frac{\sign}{\Beq} \right)^{1/2} .
\end{equation}
The product $\gaminst \tni$ for the region of the
core susceptible to interchange instability is shown in Figure $8b$ as
a function of $r/\rzero$ at the time $\delt_6$. We also plot (Fig. $8c$)
the product $\gaminst \tau_{\rm{kin}}$, where $\tau_{\rm{kin}} \equiv r/|\vn|$
is the kinematical timescale, as a function of $r/\rzero$
for the same region of the core and time. If this product is $ < 1$,
the unstable modes will be ``swept up'' by the infalling gas before they
have time to grow. It is evident from these figures that
$\gaminst \tni < 1$ and $\gaminst \tau_{\rm{kin}} > 1$ for the region of
the core susceptible to magnetic interchange. This means that there is
sufficient collisional coupling between the ions and neutrals, and that
the instability will grow before being swept along with the neutrals.
Hence, this region of the core may be interchange unstable.
An instability of this type would enhance the tansfer of gas
with a high mass-to-flux ratio to the center (e.g., Spruit \&
Taam 1990), and, as noted by Li
\& McKee (1996), might also lead to the development of
turbulence that could increase the field diffusivity in the postshock gas.
However, the onset and development of this instability can only
be studied by means of a fully 3-D simulation.
\subsection{Implications to the Magnetic Flux Problem}
The magnetic flux problem in star formation has to do with the
fact that the magnetic flux of a $1\msol$ blob of matter in the
diffuse interstellar medium
is typically several orders of magnitude greater than the flux
of a $1\msol$
protostar. Such a blob of matter would therefore have to get rid of
most of its flux before becoming a star.
Ambipolar diffusion has long been suggested as a means by which the
magnetic flux problem could be resolved (e.g., Mestel \& Spitzer 1956;
Mouschovias 1978; Paleologou \& Mouschovias 1983; Nakano 1984;
Mouschovias, Paleologou, \& Fiedler 1985). In general, these earlier
studies focused primarily on the role of ambipolar diffusion and the
magnetic flux problem for the pre-PMF phase of protostellar evolution.
\footnote{On the basis of a consideration of the timescales for
ambipolar diffusion
and Ohmic dissipation at high densities, Nakano \& Umebayashi (1986b)
suggested that significant flux loss could only take
place (primarily by Ohmic dissipation, according to their
estimates) during the dynamical phase of core
collapse. Lizano \& Shu (1989) similarly concluded that the
resolution of the protostellar magnetic flux
problem must occur during the dynamical stage of
core evolution: using the quasi-static approximation (valid for
$\nnc \simlt \rm{a~few} \times 10^4~\cc$;
Fiedler \& Mouschovias 1993; CM94; Basu \& Mouschovias 1994)
to calculate the contraction of a slightly subcritical molecular
cloud, they found that only a small amount of flux is
lost by ambipolar diffusion from the central flux tubes before
runaway collapse is initiated.}
While ambipolar diffusion does indeed reduce the flux-to-mass ratio
during that phase, the flux contained within a
$1\msol$ region of a molecular cloud core is still much larger
at the time of PMF than typical protostellar fluxes. Specifically,
CM94 and CM95 found that the
central $1\msol$ flux tube within their cores had a total
magnetic flux $\sim 10^{30}~\rm{Mx}$ (consistent with our results in
\S~3.2) during the pre-PMF dynamical collapse phase of the
typical model. This value represents a reduction by a factor
$\sim 5.6$ of the flux associated with that mass before the
onset of ambipolar diffusion. Nevertheless, it greatly exceeds
the plausible upper limit ($\sim 6 \times 10^{26}~\rm{Mx}$) on
the flux of a solar-mass protostar (estimated assuming
an average surface field of 10 kG and a stellar radius of
$10^{11}~{\rm cm}$; see Li \& McKee 1996).

We have shown in this paper that the rate of ambipolar
diffusion is strongly increased during the post-PMF epoch of star formation.
It is therefore of interest to examine the implications of our
simulation results to the magnetic flux
problem. As discussed in \S~2.2, in our calculations we only consider
the core regions at radii $r \simgt \rinner$ ($\simeq 7.3~\rm{AU}$ for
our typical model), where ambipolar diffusion is the dominant mechanism
of flux loss.
Initially, the magnetic flux contained within $\rinner$ is
$\phicent(\delt=0) = 6.7 \times 10^{26}~\rm{Mx}$. 
As shown in Figure $3a$, the central flux increases before the onset of
rapid ambipolar diffusion. This continues to the 
time $\delt \approx 10^3~\rm{yr}$.
For $\delt > 10^3~\rm{yr}$, ambipolar diffusion
prevents further advection of flux from $r > \rinner$ into the central
sink, and $\phicent$ changes very little after this time. By the time
$\delt \approx 10^5~\rm{yr}$, when $\mcent \approx 1\msol$,
$\phicent \approx 5 \times 10^{27}~\rm{Mx}$.
This represents a decrease of over two orders of magnitude relative
to the flux
associated with this mass at the time of PMF. While this value
is still about an order of magnitude higher than our adopted upper limit on the
protostelar flux, the discrepancy is now much lower than
previous estimates of ambipolar diffusion have indicated. The
important conclusion from our work is thus that {\em ambipolar
diffusion in contracting molecular cloud cores can in principle
contribute significantly to the
resolution of the magnetic flux problem} by reducing the magnetic flux
brought into a solar-mass protostar by a factor $\simgt
10^3$. The new, and somewhat surprising, result is that
{\em most of this reduction occurs after PMF.} 

The remainder of the protostellar magnetic flux could possibly
be extracted from the
infalling mass through Ohmic dissipation within $\rinner$,
although refreezing of the magnetic field
into the matter, brought about by collisional reionization at
densities $\nn \simgt 10^{14}~\cc$ (e.g.,
Pneuman \& Mitchell 1965; Nakano \& Umebayashi 1986b; Li \& McKee 1996),
as well as anomalous diffusivity
(e.g., Norman \& Heyvaerts 1985) operating in the
reionized gas, could complicate the issue. Another complicating
factor is the strong likelihood that much of the mass and flux
carried into the protostar pass through a rotationally
supported, circumstellar accretion disk of size $\gg \rinner$ (e.g., 
Lubow, Papaloizou, \& Pringle 1994; Reyes-Ruiz \& Stepinski
1996; Li 1996).
It is also conceivable that magnetic flux is
brought to the vicinity of the protostar but excluded from its
interior by turbulent diffusivity associated
with convection.
Since the region within $\rinner$ was excluded from our
calculation, we do not pursue this topic any further in this paper.
\section{Summary}
We have simulated the formation and growth of a central (i.e., protostellar)
point mass in a gravitationally collapsing core of a nonrotating magnetic
cloud modeled as an isothermal and
axisymmetric thin disk. Following up on the results of previous
simulations, we concentrated in this study on the core evolution after
point-mass formation (PMF), paying particular attention to the
role of ambipolar diffusion during this phase. In view of our assumptions
of gas isothermality and magnetic flux freezing into the
ions, our results are applicable only on scales $r \simgt 5~\rm{AU}$.

Just prior to the formation of a point mass, the model core is dynamically
collapsing (though not freely falling), with infall speeds that are
comparable to, or exceed, the isothermal speed of sound and the local
fast-magnetosonic speed, and accelerations ranging from 0.25 to 0.5 of the
local gravitational acceleration.

We have calculated the evolution of the central protostar up to the time
that it has grown in mass to $1 \msol$. Ambipolar diffusion causes the
evolution of our model core to differ significantly from that found
in previous calculations of dynamically collapsing cores, which
considered either nonmagnetic or magnetic but perfectly conducting
clouds. In particular, we find that
ambipolar diffusion in the weakly ionized gas surrounding the central
protostar is ``revitalized'' by the increase in the strength of
the gravitational field brought about by the formation and growth of
the central point mass.
(An alternative, but equivalent,
explanation of this revitalization is that the magnetic tension
force acting on the ions increases to the point where the
ion--neutral drift speed becomes comparable to the neutral
inflow speed.)
The ambipolar diffusion becomes rapid
enough to stop the inward advection of magnetic flux, which
therefore begins to pile up in the inner regions of the
collapsing core. The region of piled-up flux develops into a
hydromagnetic disturbance that propagates
outward and increases the local magnetic field amplitude away from the
axis of symmetry. As the field increases and the disturbance
reaches regions where the ion--neutral coupling becomes
stronger, the hydromagnetic front evolves into a (C-type)
shock: for the physical parameters assumed in our typical model,
this occurs at radii $r \simgt 90~\rm{AU}$.
The shock speed is supersonic (and super fast-magnetosonic) with
respect to the infalling upstream gas, but for $r \simgt
140~\rm{AU}$ it becomes subsonic in the protostellar frame.
The shock decelerates the neutrals and interrupts their infall, thereby
decreasing the accretion rate onto the central protostar: we confirmed
that the accretion rate obtained by accounting for the effect of
ambipolar diffusion was less than that calculated in a model that
evolved under the assumption of flux freezing into the
neutrals. Far behind the shock (at radii $\sim 10 - 40~\rm{AU}$
in our typical model) the neutrals are reaccelerated to free-fall, and
the column density and neutral infall speed scale as $r^{-1/2}$. The
column density scales with the time $\delt$ since PMF as $\sign
\propto (\delt)^{\sdelt}$, with $\sdelt$ lying in the range
$-0.76 \simlt \sdelt \simlt -0.55$ (whose upper bound is close to
the value $\sdelt = -0.5$ obtained under the assumption of self-similarity).
Our numerical results are in overall agreement with the simplified
analytic model of Li \& McKee (1996), who first predicted the existence
of the outward-propagating C-shock. We have found that, after PMF, this
shock is driven primarily by ambipolar diffusion, which leads to
field--matter decoupling on scales larger than those where Ohmic
diffusivity effects are important.
CCK97 derived a semianalytic similarity solution that
incorporates ambipolar diffusion and captures the main features
of the post-PMF core evolution (including the shock formation)
found in our simulation.

For the typical model presented in this paper, the protostellar
accretion rate increases from $\sim 5~\msol~\rm{Myr}^{-1}$
just prior to protostar formation to a maximum $\sim 9.4~\msol~\rm{Myr}^{-1}$ 
at $\delt \simeq 10^3~\rm{yr}$ (although, because of the
approximations involved in our calculation of
the radial magnetic field component at the inner boundary,
the actual maximum accretion rate might be higher). The
accretion rate subsequently decreases, largely on account
of the interruption of the infall by the
hydromagnetic shock, and it is equal to
$5.6~\msol~\rm{Myr}^{-1}$ by the time ($1.5 \times 10^5~\rm{yr}$ after
PMF) the central mass has grown to $1~\msol$.
For comparison, the ``canonical'' accretion rate ($\simeq
C^3/G$; Shu 1977) for this model is
$1.5~\msol~\rm{Myr}^{-1}$, although it has been recognized
that this value would be larger in clouds where magnetic
stresses supplement thermal pressure support.
[For example, Li
\& Shu (1997) estimated an increase by a factor $\sim (1 + H_0)$,
where $H_0$ ($\sim 1$) is the magnetically supported fractional
overdensity in the stationary, pre-PMF cloud.
Other magnetic models (e.g., Tomisaka 1996; Basu 1997) have also indicated a
larger-than-canonical average accretion rate.]
The decline in
the accretion rate with time
is consistent with the trend inferred from observations of Class 0 and
Class I sources. Furthermore, the accretion rate and the
circumstellar mass at radii $\simgt 500~\rm{AU}$
in our typical model are comparable to those
derived by millimeter observations of protostars with ``infalling''
disks, such as HL Tauri and L1551-IRS5. The magnetic field structure
in our model is consistent with polarimetric observations
of molecular clouds that reveal essentially uniform fields in cloud
envelopes ($r \simgt 0.1~\rm{pc}$), and with sub-mm polarimetric surveys
that indicate hourglass field structures within cloud cores.

Ambipolar diffusion is so effective after PMF that it leads to a
reduction by more than two orders of magnitude in the flux
threading the central $1~\msol$, compared with less than one
order of magnitude reduction in the flux threading this mass
between the start of the cloud contraction and the time of
PMF. Altogether, ambipolar diffusion reduces the flux by more
than three orders of magnitude and brings it to within an order
of magnitude of the estimated upper limit on the flux of a $\sim
1~\msol$ protostar. Ambipolar diffusion in collapsing cloud cores
could thus go a long way toward resolving the
protostellar magnetic flux problem, although to fully address
this issue one would need to consider Ohmic diffusivity on
scales smaller than those included in the present calculation as
well as the effects of rotation.

Large ($\simgt 1~{\rm{km}}~{\rm s}^{-1}$) ion--neutral drift
speeds occur in our representative simulation on scales $r
\simlt 200~\rm{AU}$ for $\delt \simgt 2 \times
10^4~\rm{yr}$, and are a unique prediction of our model.
Observational detection of this effect could be used to
distinguish our results from other (nonmagnetic and magnetic) collapse
models that do not account for ambipolar diffusion. We have also
confirmed that the postshock region is susceptible to the
magnetic interchange instability, as first suggested by Li \& McKee
(1996). This instability could only develop outside the
region of large ion--neutral drift velocities, where it is not
subject to quenching by ambipolar diffusion effects. Magnetic interchange
would increase the rate of mass transfer to the center and could
also lead to turbulence that might enhance the magnetic
diffusivity. A numerical investigation of this instability
does, however, require a 3-D simulation.

The mass and magnetic flux distributions during the post-PMF epoch are
directly relevant to the question of protostellar disk formation and
the generation of disk outflows by magnetic ejection mechanisms
(e.g., K\"{o}nigl \& Ruden 1993). However, in order to model the
formation of centrifugally supported circumstellar disks, it
is necessary to incorporate rotation and magnetic braking into
the core-evolution calculations. This will be considered in a future
publication.

\acknowledgements{This work was supported in part by NASA grants NAG 5-2266
and NAG 5-3687. We thank the referee, T. Hartquist, for several
suggestions that helped improve the presentation of this paper.
A discussion with R. Hildebrand on polarimetric observations of ordered
magnetic fields in and around interstellar clouds is gratefully
appreciated.
We also acknowledge useful input from C. McKee, Z.-Y. Li, P. Andr\'{e},
and F. Shu.
GC would especially like to thank T. Mouschovias and S. Morton for early
discussions on the nature and formulation of the problem presented in
this paper.
}
\setcounter{equation}{0}
\\ \\
\renewcommand{\theequation}{A{\arabic{equation}}}
\bc {\bf APPENDIX A} \ec
\bc {\bf DESCRIPTION OF MODEL PARAMETERS} \ec

We consider isothermal,
magnetic interstellar molecular clouds consisting of neutral molecules
($\HII$ with a 20\% He abundance), singly charged molecular and atomic
ions (such as $\HCOp$, $\Mgp$, $\Nap$, $\Fep$, $\Cp$, $\Sup$, and
$\Sip$), and electrons. For the purposes of an accurate calculation of
the equilibrium abundances of charged particles, we also include (negative)
singly charged grains and neutral grains. However, the collisional effects of
grains on the neutrals (which in certain cases can be significant; see
CM94, CM95) are ignored in this paper: this is done by setting the
parameter $\siggn$ (see below) equal to 0. The effects of magnetic
braking and rotation, which can also affect the evolution of a core
(Basu \& Mouschovias 1994, 1995a, b) are similarly neglected in this paper.
They will be accounted for in a later publication.

The assumptions that model clouds are thin and that evolution 
along field lines is quasistatic allow the time-dependent
nonlinear set of nonideal MHD equations that govern the evolution of
the model cloud to be integrated over $z$, thus reducing the
dimensionality of the formal problem (CM93). The simplified set of
equations is listed as equations (A1)--(A17) in CM95. The equations are
cast in dimensionless form by adopting the quantity $2 \pi G \sigcref$
as the unit of acceleration, $\Bref$ as the magnetic field strength,
and the isothermal speed of sound $C$ as the unit of velocity. The
quantity $\sigcref$
is the central column density of a reference state used to specify how
magnetic field lines are initially loaded with mass, and $\Bref$,
as mentioned in \S~2.2, is the background magnetic field strength
at infinity. The implied units of time, length, and mass density are,
respectively, $C/2 \pi G \sigcref$, $C^2/2 \pi G \sigcref$, and
$2 \pi G \sigcref^2/C^2$. The resulting system of dimensionless
equations contains the following set of nondimensional parameters:
\begin{eqnarray}
\label{mtohphieq}
\phitomd &\equiv& \frac{\left(d M/d \phiB\right)_{\rm{c0}}}{\left(d M /d \phiB\right)_{\rm{d,crit}}}
= \frac{\left(\sigcref/\Bref\right)}{\left(1/2 \pi \sqrt{G}\right)}
= 0.196 \left(\frac{\sigcref}{3.63 \times 10^{-3}~{\rm g}~{\rm{cm}}^{-2}}\right) \left(\frac{30~\mu{\rm G}}{\Bref}\right), \hspace{2em} \\
\label{Pextdeq}
\Pextd &\equiv& \frac{\Pext}{\left(\pi/2\right)G \sigcref^2}
=0.1 \left(\frac{\Pext}{1.38 \times 10^{-13}~{\rm{dyn}}~{\rm{cm}}^{-2}}\right) \left(\frac{3.63 \times 10^{-3}~{\rm g}~{\rm{cm}}^{-2}}{\sigcref}\right), \\
\label{sigHIIeq}
\sigHII &\equiv& \frac{\sigHI}{1.4 \left[1 + \left(\mHII/\mi\right)\right]} \frac{\left( 2 \pi G \sigcref \right)^2}{C^5} \\ \nonumber 
&=&1.2 \times 10^{-48} \frac{\left(\sigHI/1.7 \times 10^{-9}~{\rm{cm^3}}~{\rm s}^{-1}\right)}
{1 + \left[15\left(\mi/\mHII\right)\right]^{-1}}
\left(\frac{\sigcref}{3.63 \times 10^{-3}~{\rm g}~{\rm{cm}}^{-2}}\right)^2
\left(\frac{0.188~{\rm{km}}~{\rm s}^{-1}}{C}\right)^5 , \hspace{2em}
\\
\label{siggneq}
\siggn &\equiv& \siggna \frac{\left(2 \pi G \sigcref \right)^2}{C^5} 
\nonumber\\ 
&=& 9.3 \times 10^{-47} \left(\frac{a}{10^{-6}~{\rm{cm}}}\right)^2
\left(\frac{\sigcref}{3.63 \times 10^{-3}~{\rm g}~{\rm{cm}}^{-2}}\right)^2
\left(\frac{0.188~{\rm{km}}~{\rm s}^{-1}}{C}\right)^4, 
\end{eqnarray}
where the quantity $a$ in equation (\ref{siggneq}) is the grain radius.
The parameter $\phitomd$ is the initial central mass-to-flux ratio
in units of the critical value for collapse. $\Pextd$ is the constant
external pressure normalized to the self-gravitational stress of the
matter contained within the central flux tube of the reference state.
The parameter $\sigHII$ is the dimensionless neutral--ion collisional
rate, and $\siggn$ is the dimensionless neutral--grain collisional
rate. The initial equilibrium state introduces one more parameter,
the quantity $\lref$, a length scale in the column density of the
reference state (see CM93, eqs. [60b] and [73b]). The dimensionless
cloud radius is taken to be $5 \, \lref$ in this paper. A final dimensionless
parameter is the initial dust-to-gas mass ratio, $\chi_{\rm{g,0}}$,
which is equal to 0.01 in the typical model.

Abundances of charged particles are determined by solution of the chemical
rate equations with balance between creation and destruction of the
various charged species specified above, accounting for such processes as
cosmic-ray ionization, dissociative recombination of molecular ions and
electrons, radiative recombination of atomic ions and electrons,
attachment of charged particles on grains, and charge transfer between
molecular and atomic ions; we also account for ionization due to an
external (interstellar) ultraviolet radiation field. The relevant
chemical equations are given by equations (7)--(17) of CM95. They
contain six relevant nondimensional parameters of the form
$\zeta_{{}_{\alpha_{0},\rm{UV,CR}}} \equiv \zeta_{{}_{\alpha_{0},\rm{UV}}}/\zcr$,
where $\zeta_{\alpha_0}$ is the UV ionization rate at the cloud boundary
for neutral species $\alpha_0$, and $\zcr$ is the cosmic-ray ionization
rate.
\bc {\bf APPENDIX B} \ec
\bc {\bf DENSITY SCALING IN THE INNER FLUX TUBES OF MODEL CLOUDS} \ec

For our thin-disk model, the density $\rhon$ ($=\mn \nn$) is calculated
from equation (\ref{newforcbaleq}). We examine the power-law behavior
of the density in the inner flux tubes of the core for the case in
which the tidal gravitational field of the central protostar is negligible,
and then for the opposite case where it is the dominant term in
equation (\ref{newforcbaleq}). In the first case, using the fact
that $\Pext \ll (\pi/2) G \sign^2$ in the inner flux tubes of
the core
and that the magnetic squeezing term is also negligible
there for most times of interest (see discussion in \S 2.2
following eq. [\ref{newforcbaleq}]), the above equation
yields $\rhon = \pi G \sign^2/2 C^2$.
During the later stages of the pre-PMF phase of core contraction,
$\sign \propto r^{-1}$ (CM94, CM95; Basu \& Mouschovias 1994, 1995a,b;
see also \S~3.2). Therefore $\rhon \propto r^{-2}$ as the core
approaches PMF.

We now turn to the case where the tidal stress from the central protostar
is the dominant term on the r.h.s. of equation (\ref{newforcbaleq}).
Near the protostar, tidal squeezing of the disk (see Fig. $6i$) results
in $Z/r = \sign/2 \rhon r \ll 1$.
\footnote{This squeezing will be augmented,
particularly at late times, by magnetic field effects
represented by the second term on the r.h.s. of
equation (\ref{newforcbaleq}).}
Expansion of the second term in braces
on the r.h.s. of equation (\ref{newforcbaleq})
gives the solution $\rhon = (G \mcent/8)^{1/2} \sign/C r^{3/2}$. In the
limit of free-fall collapse near the protostar $\sign \propto r^{-1/2}$
(see \S~3.3) and $\rhon$ is again $\propto r^{-2}$. Hence, the power-law
behavior of the density in our model is constrained to be
$\propto r^{-2}$ before and after PMF.
This result is different from the behavior $\rhon \propto r^{-3/2}$,
which would occur for the case of spherical infall in the gravitational
field of a central point mass. This difference is due to our use of
the vertical one-zone approximation for the mass density 
[$\rhon(z,r,t)=\rhon(0,r,t) \equiv \rhon(r,t)$] in equation (\ref{newforcbaleq})
for balance of forces along field lines. Figure $9$ shows the number
density $\nn$, normalized to $\nncb = 2.6 \times 10^3~\cc$, as
a function of $r/\rzero$ at the same seven times $\delt_j$ as in Figure
6. The $r^{-2}$ scaling of the density is apparent in this
figure at small radii.
%
\bc {\bf APPENDIX C} \ec
\renewcommand{\theequation}{C{\arabic{equation}}}
\bc {\bf BREAKDOWN OF FLUX FREEZING AFTER POINT-MASS FORMATION} \ec

In this Appendix we present a heuristic argument that
demonstrates the breakdown of flux freezing, and the corresponding
revitalization of ambipolar diffusion, following point-mass
formation (PMF). We show this by means of a {\em reductio ad
absurdum} argument: we {\em assume} that the collapse proceeds
under flux freezing conditions, and we use the frozen-flux
solution to evaluate the ratio of the ion--neutral drift speed
($\vd$) and the free-fall speed ($\vff = [2 G M(r)/r]^{1/2}$).
Flux freezing corresponds to $\vd \ll \vff$; however, by
deriving the magnetic force on the ions
from the numerically computed evolution and using the ion equation
of motion to calculate the resulting ion--neutral drift speed, we
find that the ratio $\vd/\vff$ becomes $\gg 1$ after PMF. This implies that
the assumption of flux freezing is {\em not self-consistent} and that
ambipolar diffusion {\em necessarily} sets in. According to this
argument, the increase of $\vd/\vff$ in the inner core
following PMF can be
attributed to the growth of the magnetic tension term in the ion
force equation. The quantity $\vd/\vff$ is essentially the inverse of the
ratio of the ambipolar diffusion
timescale ($\tad \equiv r/\vd$) and gravitational contraction timescale
[$\tgr \equiv (r/|\gr|)^{1/2} \approx \tff$] considered in
\S~3.3 (see eq. [\ref{tadeq}]). In the
argument presented in \S~3.3 (see also \S~1), $\tad/\tgr$ is
shown to decrease below 1 during that epoch on account of the effect of
the central point mass on the magnitude of the gravitational
acceleration in its
vicinity. As we show below, these two alternative
descriptions of the ambipolar-diffusion revitalization process
are equivalent.

In a disk-like cloud, the dominant terms in the expression for the
radial magnetic force per unit mass are
\begin{eqnarray}
\label{Fmageq}
\frac{F_{{\rm{mag}},r}}{\sign} &=& \frac{1}{2 \pi \sign} \left(\Beq \BrZ
- Z \frac{\partial \Beq}{\partial r}\right),  \nonumber \\
&=& \frac{\Beq \BrZ}{2 \pi \sign}\left(1 - \frac{Z}{\Beq}\frac{\partial \Beq}{\partial r}\right) 
\end{eqnarray}
(see eq. [28c] of Ciolek \& Mouschovias 1993). The first term on the
r.h.s. of equation (\ref{Fmageq}) is the magnetic tension force,
while the second term is the magnetic pressure force.
It turns out (see below) that, in a frozen-flux core, the tension term
comes to dominate after PMF. In that case, using the ion force
equation
\begin{equation}
\label{ionforceqb}
\frac{\sign}{\tni} \vd = F_{{\rm{mag}},r}
\end{equation}
and the continuity equation
($\mdot = - 2 \pi r \sign \vn$), and assuming $|\vn| \approx \vff$ in the
inner flux tubes of a collapsing core, one obtains
\begin{equation}
\label{vrateq}
\frac{\vd}{\vff} = \left[\frac{\tni}{2 \mu_{B}^2}
\frac{\mdot}{M}\right] \frac{\BrZ}{\Beq} \ , 
\end{equation}
where $\mu_B$ is the mass-to-flux ratio in units of the critical
value for collapse, as discussed in \S~1. The term in brackets on the
r.h.s. of equation (\ref{vrateq}) is the same as equation (4)
of Li \& McKee (1996), who used $\epsilon = \mu_{B}^{-1}$ in their expression.
Consider now the behavior of $\BrZ$ and $\Beq$ in a frozen-flux
core following PMF. As was noted in \S~2.2, $\Beq \propto \sign$ and
$\BrZ \propto \gr$ in a thin, perfectly conducting disk with $\mu_B =
const$. Thus, outside the central sink cell, $\BrZ \approx
\phiB(r,t)/2 \pi r^2 \approx \phicent/2 \pi r^2$ after PMF (see eq.
[\ref{newbreq}] and associated discussion). Here $\phiB(r,t)$ is the
total flux enclosed within radius $r$, which is $\sim \phicent$ after PMF.
Hence, $\BrZ$ scales as $r^{-2}$. Near the central point mass the
inflowing matter is in approximate free fall and hence
$\Beq \propto \sign \propto r^{-1/2}$ (see \S~3.3). Therefore, after
PMF, $\BrZ/\Beq \propto \phicent/r^{3/2}$ will become $\gg 1$ as
$r \rightarrow 0$. This behavior has indeed been seen in our frozen-flux
model simulations, and is also apparent in the self-similar
frozen-field collapse calculations of Li \& Shu (1997) and CCK. 

Because of the increase of $\BrZ/\Beq$ near the origin after
PMF, the ratio $\vd/\vff$ given by equation (\ref{vrateq}) will also
become $\gg 1$ there. This is illustrated graphically in
Figures $10a$, $10b$, and $10c$, which show, respectively, the
ratios $\vd/|\vn|$, $\BrZ/\Beq$, and $F_{{\rm{mag}},r}/F_{\rm{mag,pres}}$ in
the inner flux tubes of the frozen-flux model core presented 
in \S~3.3. The quantity $F_{{\rm{mag}},r}$ is the actual magnetic force per
unit mass calculated in our model (given by eq. [\ref{Fmageq}]
above), which includes magnetic tension, whereas
$F_{\rm{mag,pres}} = \Beq^2/2 \pi \sign$
is the force term used by Li \& McKee (1996), which
approximately represents the effect of magnetic pressure
alone. It is seen from these figures
that $\vd/|\vn|$ does indeed become $\gg 1$ as the core evolves,
and that this is due to the
increase of $\BrZ/\Beq$, which scales as $r^{-3/2}$, as expected.
(As pointed out in \S~3.3, our frozen-flux model is calculated
by setting $\vi=\vn$, so that $\vd=0$ in our numerical code;
to obtain $\vd$ in Fig. $10a$, we used equations
[\ref{ionforceqb}] and [\ref{Fmageq}].) These
results also confirm that $F_{{\rm{mag}},r}/F_{\rm{mag,pres}} = \BrZ/\Beq$.
The fact that $\vd/|\vn|$ becomes $\gg 1$ {\em means that the
assumption of continued flux freezing is not self-consistent and that
ambipolar diffusion must eventually take place}.
\footnote{This conclusion
did not follow from the estimate of $\vd/\vff$ in Li \& McKee
(1996) because they took the radial magnetic force per unit mass
to be $F_{\rm{mag,pres}}$ rather than $F_{{\rm{mag}},r}$ and thus
omitted the ratio $\BrZ/\Beq$ from their equation (4). The
argument presented in this Appendix also explains why the
revitalization of ambipolar diffusion following PMF cannot be
studied using a spherically symmetric model (e.g., Li 1998),
where only the magnetic pressure force, but not the tension
force, is taken into account.}
The region of flux
decoupling will move outward with time, which is consistent with
the fact that ambipolar diffusion in our simulations first becomes
noticeable on our smallest resolvable scale, and that the decoupling front
subsequently moves outward.

This ``$\BrZ/\Beq$'' explanation of the onset of ambipolar
diffusion is the same as the ``$\tgr/\tni$'' argument
given in \S\S~1 and 3.3, where we emphasized the
fact that $\tad/\tgr \sim \tgr/\tni$. This can be seen in the
following way. Substituting $\BrZ \approx \phiB(r,t)/2 \pi r^2$
into the ion force equation (\ref{ionforceqb}), we get
\begin{equation}
\label{vdeq}
\vd = \frac{\tni}{2 \pi \sign} \Beq \frac{\phiB}{2 \pi r^2} \ .
\end{equation}
Dividing this equation by $\vff$, and using the relation
$M = \vff^2 r/2 G$, yields
\begin{equation}
\label{vratbeq}
\frac{\vd}{\vff} = \tni \left[ \frac{\Beq}{4 \pi^2 G \sign}
\frac{\phiB}{M} \right] \frac{\vff}{2r} = \frac{1}{\sqrt{2} \mu_{B}^2}
\frac{\tni}{\tgr} ~,
\end{equation}
where, in the last equality, we used $\tgr = \sqrt{2} r/\vff$ and
the definition of $\mu_B$. From this equation it follows that
$\tad/\tgr \approx \vff/\vd = \mu_{B}^2 \tgr/\tni$. The expression given
by equation (\ref{vratbeq}) is the same as that derived by Mouschovias (1991,
eq. [10a]), who found that $\tad \propto (\tgr^2/\tni)
(\Phi_{B,\rm{crit}}/\phiB)^2$,
where $\Phi_{B,\rm{crit}}$ is the critical magnetic flux.
Finally, by inserting equations (\ref{ionforceqb}) and (\ref{vdeq}) into equation
(\ref{rforceeq}) and using the relations $|\gr| \approx G M/r^2$
and $\BrZ \approx \phiB/2 \pi r^2$
in the central flux tubes of a core following PMF (see eqs. [\ref{newgreq}]
and [\ref{newbreq}]), one obtains (neglecting thermal-pressure forces)
\begin{equation}
\label{arateq}
\left(1 - \frac{|\an|}{|\gr|}\right) = \left(1 - \frac{\tgr^2}{\tacc^2}\right) = \frac{\left(\Beq/\sign\right) \left(\phiB/M\right)}{4 \pi^2 G} = \mu_{B}^{-2}~. 
\end{equation}
(This relation can also be derived from eq. [24] of Basu 1997.)
Therefore, equations (\ref{vratbeq}) and (\ref{arateq}) yield
equation (\ref{tadeq}), and the ``$\BrZ/\Beq$" and ``$\tgr/\tni$"
pictures are seen to be completely equivalent, as claimed.
\newpage
\bc {\bf {\Large {\sc Table 1}}} \\ 
{\large \sc Dimensionless Parameters of a Typical Model Cloud}${}^\dagger$ \\ 
\ec
\bc
\begin{tabular}{llllr} \hline\hline
\multicolumn{1}{c}{${\mu_{\rm{d,c0}}}$}
&\multicolumn{1}{c}{\hspace{1.7em}$\lref$}
&\multicolumn{1}{c}{\hspace{2em}$\Pextd$}
&\multicolumn{1}{c}{\hspace{2em}$\sigHII$}
&\multicolumn{1}{c}{\hspace{2em}$\siggn$}
\\
\hline
\multicolumn{1}{c}{0.256}
&\multicolumn{1}{c}{\hspace{2em}5.5$\pi$}
&\multicolumn{1}{c}{\hspace{2em}0.1}
&\multicolumn{1}{c}{\hspace{2em}$2.53 \times 10^{-48}$}
&\multicolumn{1}{c}{\hspace{2em}0} \\
\hline
\hline
\end{tabular}
\ec
${}^\dagger$ For a discussion of the meaning of these parameters,
see Appendix A.
\bc {\bf {\Large {\sc Table 2}}} \\ 
{\large \sc Dimensionless UV Ionization Parameters ${}^{\dagger \dagger}$} \\ 
\ec
\bc
\begin{tabular}{lllllr} \hline\hline
\multicolumn{1}{c}{$\zC$}
&\multicolumn{1}{c}{\hspace{2em}$\zMg$}
&\multicolumn{1}{c}{\hspace{2em}$\zNa$}
&\multicolumn{1}{c}{\hspace{2em}$\zFe$}
&\multicolumn{1}{c}{\hspace{2em}$\zS$}
&\multicolumn{1}{c}{\hspace{2em}$\zSi$}
\\
\multicolumn{1}{c}{$(10^6)$}
&\multicolumn{1}{c}{\hspace{2em}$(10^6)$}
&\multicolumn{1}{c}{\hspace{2em}$(10^6)$}
&\multicolumn{1}{c}{\hspace{2em}$(10^6)$}
&\multicolumn{1}{c}{\hspace{2em}$(10^6)$}
&\multicolumn{1}{c}{\hspace{2em}$(10^6)$} \\
\hline
\multicolumn{1}{c}{$2.60$}
&\multicolumn{1}{c}{\hspace{2em}$0.90$}
&\multicolumn{1}{c}{\hspace{2em}$0.11$}
&\multicolumn{1}{c}{\hspace{2em}$2.40$}
&\multicolumn{1}{c}{\hspace{2em}$14.4$}
&\multicolumn{1}{c}{\hspace{2em}$24.0$}
\\
\hline
\hline
\end{tabular}
\ec
${}^{\dagger \dagger}$ All ionization rates are normalized to a cosmic-ray
ionization rate of $5 \times 10^{-17}~{\rm s}^{-1}$. All dimensional
UV ionization rates are taken from Table 9 of Black \& Dalgarno (1977,
and references therein).

\newpage
\bc {\bf Figure Captions} \ec
\newcounter{figcount}
\begin{list}
{Fig. \arabic{figcount}.$-$}{\usecounter{figcount}}
            \setlength{\rightmargin}{\leftmargin} 
\item {\it Spatial profiles of physical quantities in the typical model,
prior to point-mass formation,
as functions of radius $r$ (normalized to the initial cloud radius
$R_{0}$
$=4.29~\rm{pc}$) at eleven different times $t_j$ ($j=0,1,2,...,10$)
chosen such that the central density has increased by a factor of $10^j$
with respect to its initial value}. These times are $0$, $7.56~\rm{Myr}$,
$9.29~\rm{Myr}$, $9.56~\rm{Myr}$, $9.607~\rm{Myr}$, $9.6167~\rm{Myr}$,
$9.6169~\rm{Myr}$, $9.6190~\rm{Myr}$, $9.61977~\rm{Myr}$,
$9.61981~\rm{Myr}$, and $9.61983~\rm{Myr}$, respectively. An {\it asterisk}
on a curve, present only when a supercritical core has formed, marks the
instantaneous position of the critical flux tube. An {\it open circle}
on every curve locates the instantaneous position of the critical
thermal lengthscale. As discussed in \S~2.2,
the assumption of isothermality and the form of the
induction equation used in our calculation are not valid for
densities $\nn > 10^{11}~{\cc} \approx 4 \times 10^7~\nncb$.
({\it a}) Neutral density, normalized to its initial
central value $\nncb$ ($=2.6 \times 10^3~\cc$). ({\it b}) Neutral
column density, normalized to its initial central value $\sigcb$ ($=5.59 \times
10^{-3}~{\rm g}~\rm{cm}^{-2}$).
({\it c}) Mass-to-flux ratio, in units of the critical value for
gravitational collapse. Between times $t_1$ and $t_2$ a critical core
has formed (because of ambipolar diffusion) inside a magnetically
subcritical envelope. ({\it d}) Vertical ($z$) component of the magnetic
field in the equatorial plane of the disk, normalized to its initial
central value $\Beqcb$ ($=35.3~\mu{\rm G}$). ({\it e}) Radial ($r$) component
of the magnetic field at the upper surface of the cloud, normalized
to $\Beqcb$. ({\it f}) Infall speed of the neutrals, normalized
to the isothermal speed of sound $C$ ($=0.19~\rm{km}~\rm{s}^{-1}$).
By the end of the run, $|\vn| \approx C$ in the inner core.
({\it g}) Ion--neutral drift speed ($\equiv \vi - \vn$),
normalized as in ({\it f}). Ion depletion at higher densities is
responsible for the appearance of the maximum in the core for $t \geq t_6$.
({\it h}) Mass infall rate, in units of $\msol~\rm{Myr}^{-1}$.
({\it i}) Ratio of cloud vertical half-thickness $Z(r)$ and radius $r$.
\item{{\it Accretion in the central cell during the PMF epoch.}
({\it a}) Central (protostellar) mass, in $\msol$, as a function
of time $\delt$ ($=t-t_{10}$). Formation of a point mass occurs at
$\delt \simlt 200~\rm{yr}$. ({\it b}) Accretion rate in $\msol~\rm{Myr}^{-1}$.
{\it Dashed} curve refers to a model that did not include the effect
of ambipolar diffusion for $\delt \geq 0$. For $\delt \simgt 4 \times 10^3~\rm{yr}$,
ambipolar diffusion, through the formation
of a hydromagnetic shock in the neutrals, decreases the protostellar
accretion rate of the typical model. ({\it c}) Accretion rate
in $\msol~\rm{Myr}^{-1}$ as a function of central mass (in $\msol$).
{\it Dashed} curve refers to the frozen-flux model, as in ({\it b}). 
\item {\it Physical quantities in two different computational mesh
cells, as functions of $\delt$ ($\approx$ the time elapsed since PMF).}
({\it a}) Cell 1: ratio of ambipolar-diffusion timescale $\tad$
and gravitational contraction timescale $\tgr$, ratio of
gravitational timescale and neutral--ion collision timescale $\tni$, and
(normalized) magnetic flux $\phiB$ contained within the cell.
The gravitational contraction timescale
decreases because of an increase in the magnitude of the gravitational
field associated with the central accreting protostar, which
increases the rate of
ambipolar diffusion in the inner flux tubes of the model cloud.
For $\delt \simgt 10^3~\rm{yr}$, ambipolar diffusion has become rapid
enough to halt the growth of magnetic flux
within this cell. ({\it b}) Same as ({\it a}), but for cell 5. The
evolution is similar to that of cell 1
except that it exhibits a time delay,
indicating that the region of piled-up magnetic flux is advancing
outward from the center.
\item {\it Physical quantities in computational mesh cells $l=2, 5,
10, 15, 20, 25$, and $27$, as functions of $\delt$.}
({\it a}) Vertical ($z$) component of the magnetic field in the
equatorial plane of the disk, normalized
to $\Beqcb$. Prior to the passage of the hydromagnetic
disturbance (HMD),
the magnetic field strength decreases because of inward advection of
magnetic flux. At later times the magnetic field is increased by
the
compression induced by the outward-propagating HMD.
({\it b}) Column density, normalized to $\sigcb$. In cells 2 and
5 the inflow of the neutrals is unaffected by the advancement of
the HMD, and the column density decreases with time
due to advection of matter into the central sink. In cells $l > 8$ the
combination of the enhanced field strength (due to the piled-up magnetic flux)
and the increased collisional coupling between the ions and neutrals
at larger radii eventually leads to a deceleration of the
infalling neutrals and the formation of a hydromagnetic shock. The column
density increases immediately after the passage of the shock front;
at later times the neutrals diffuse through the shock and are reaccelerated
to free fall,
with $\sign \propto (\delt)^{\sdelt}$, $ -0.76 \simlt \sdelt \simlt -0.55$.
\item {\it Outward motion of the hydromagnetic disturbance as a
function of $\delt$}. ({\it a}) Instantaneous
position of the HMD, normalized to the cloud radius $\rzero$.
({\it b}) Speed of the disturbance (in the protostellar rest frame),
normalized to $C$. Early on, the expansion of the HMD is nonsteady. At
later times, the speed of the disturbance is subsonic. ({\it c}) Mach
number relative to the infalling neutrals just upstream of the
disturbance. Note that the time intervals between successive
data points in this figure are larger than those in Figs. 2--4, reflecting
a time-averaging of the motion of the HMD. For this reason the
steady oscillations seen in Figs. 2--4 at later times do not appear
in this figure.
\item {\it Spatial profiles of physical quantities in the typical
model during the post-PMF epoch as functions of $r/\rzero$.} Seven different
times are plotted: $\delt= 0$, $2.17 \times 10^2~\rm{yr}$,
$6.11 \times 10^2~\rm{yr}$, $1.54 \times 10^3~\rm{yr}$,
$3.81 \times 10^3~\rm{yr}$, $2.38 \times 10^4~\rm{yr}$,
and $1.48 \times 10^5~\rm{yr}$. An {\it asterisk} on a curve locates,
as in Fig. 1, the instantaneous position of the critical
magnetic flux tube. ({\it a}) Mass, in $\msol$.
The protostar first forms at $\delt \approx 0$.
({\it b}) Vertical
component of the magnetic field in the equatorial plane, normalized
to $\Beqcb$. The {\it dashed} curve refers to a model in which the magnetic
flux was assumed to be frozen into the neutrals at all times after PMF,
shown at the last plotted time ($\delt_6$). The
field strength is strongly amplified behind the expanding HMD
for $\delt > \delt_2$. For $\delt_5 \simlt \delt \simlt \delt_6$
the field strength at $r/\rzero \simlt 2.5 \times 10^{-4}$
is significantly reduced,
reflecting the refreezing of magnetic flux into the
infalling neutrals brought about by the greatly enhanced collisional
coupling that occurs when $\vd > \vdcrit$.
({\it c}) Neutral column density, normalized to $\sigcb$. The shock in
the neutrals manifests itself by the enhancement and local
maximum in the
column density for $\delt \geq \delt_4$.
Far behind the shock the neutrals establish free-fall collapse, and
$\sign \propto r^{-1/2}$ for $r \simlt 4 \times 10^{-5} \rzero$. The
{\em dashed} curve refers to the model in which magnetic flux was
frozen into the neutrals, at the same time as in ($b$). 
({\it d}) Radial component of the magnetic field at the surface of the
cloud, normalized to $\Beqcb$. For $\delt > \delt_2$, ambipolar
diffusion halts the inward advection of flux and further bending of
the field lines. The {\em dashed} curve refers to the frozen-flux
model, at the same time as in ($b$). Continual inward advection of
magnetic flux in the frozen-in model results in much greater bending
of field lines than in the ambipolar-diffusion case.
({\it e}) Infall speed of the neutrals, normalized to $C$. The location
of the hydromagnetic shock is marked by the abrupt transition from
supersonic to subsonic inflow for $\delt > \delt_4$.
Free-fall collapse is established far behind the shock, and
$\vn \propto r^{-1/2}$ for $r \simlt 4 \times 10^{-5} \rzero = 35.5~\rm{AU}$.
({\it f}) Ion--neutral drift speed, normalized as
in ($e$). In the region of rapid ambipolar diffusion behind the expanding
hydromagnetic disturbance, $\vd \approx |\vn|$.
({\it g}) Accretion rate, in $\msol~\rm{Myr}^{-1}$.
For $\delt \simgt \delt_4$ the accretion rate is decreased behind
the shock front. ({\it h}) Ratio of neutral--ion collisional timescale
and gravitational contraction timescale. Near the inner boundary
this ratio is $\gg 1$, indicating efficient ambipolar diffusion.
Further away from the center, the degree of ionization increases
and the magnitude of the gravitational field decreases, resulting
in $\tni/\tgr < 1$ and, therefore, a much better collisional coupling
between the ion and neutral fluids. ({\it i}) Ratio of cloud
vertical half-thickness $Z(r)$ and radius $r$. By the time $\delt_6$
the tidal gravitational field of the central protostar has compressed
the inner core to a disk that is geometrically thin.
[Magnetic squeezing of the disk, not included in this
calculation, further reduces $Z(r)$.]
\item{\it Mass of the gas surrounding the central protostar in the
typical model, in $\msol$, as a function of $r/\rzero$.} All labels
and times $\delt_j$ are the same as in Fig. 6.
\item {\it Spatial profiles of quantities related to the stability
of the core of the typical model cloud with respect to magnetic interchange, as a function of
$r/\rzero$}.
All labels and times $\delt_j$ are the same as in Fig. 6. ({\it a})
Local mass-to-flux ratio, in
units of the critical value for collapse. Unloading of mass from field
lines due to rapid ambipolar diffusion behind the hydromagnetic
disturbance results in positive values of $d \left(\ln \sign/\Beq \right)/dr$
inside the core at times $\delt > \delt_1$. ({\it b})
Product of the linear instability growth rate and the
neutral--ion collision time at $\delt = \delt_6$ for the region of
the core susceptible to
magnetic interchange. Collisional coupling of the ions and neutrals
is sufficient to allow the instability to grow in this
region of the core. ({\it c}) Product of the linear instability
growth rate and the kinematical timescale ($= r/|\vn|$) for the
same time and region of the core as in ({\it b}). An unstable
mode can grow before being advected downstream by the fluid.
\item {\it Density profile of the typical model, normalized to $\nncb$,
as a function of $r/R_0$.} All labels and times $\delt_j$ are
the same as in Fig. 6. In the inner core, $\nn \propto r^{-2}$
throughout the post-PMF epoch.
\item {\it Physical quantities in the inner flux tubes of
the frozen-flux model as functions of $r/R_0$ at seven different
times $\delt_j$.} ({\it a}) Ratio
of ion--neutral drift speed and the neutral infall
speed. In this model, $\vi$ is set equal to $\vn$ in the numerical
code, and the drift speed $\vd$ is calculated from the ion force equation.
As the core evolves, $\vd/|\vn|$ becomes $\gg 1$, indicating that
the freezing of magnetic flux in the neutrals would break down.
({\it b}) Ratio of the $r$ and $z$ components of the magnetic field.
({\it c}) Ratio of the total radial magnetic force and the magnetic pressure
force. After PMF, the ratios $\BrZ/\Beq$ and $F_{\rm{mag},r}/F_{\rm{mag,pres}}$
scale as $r^{-3/2}$. The enhanced magnetic tension force following PMF
would act to revitalize ambipolar diffusion during this phase of the
collapse, which is indeed what is found to occur in the typical model
presented in \S~3.3.}
\end{list}
\begin{figure}
\plotone{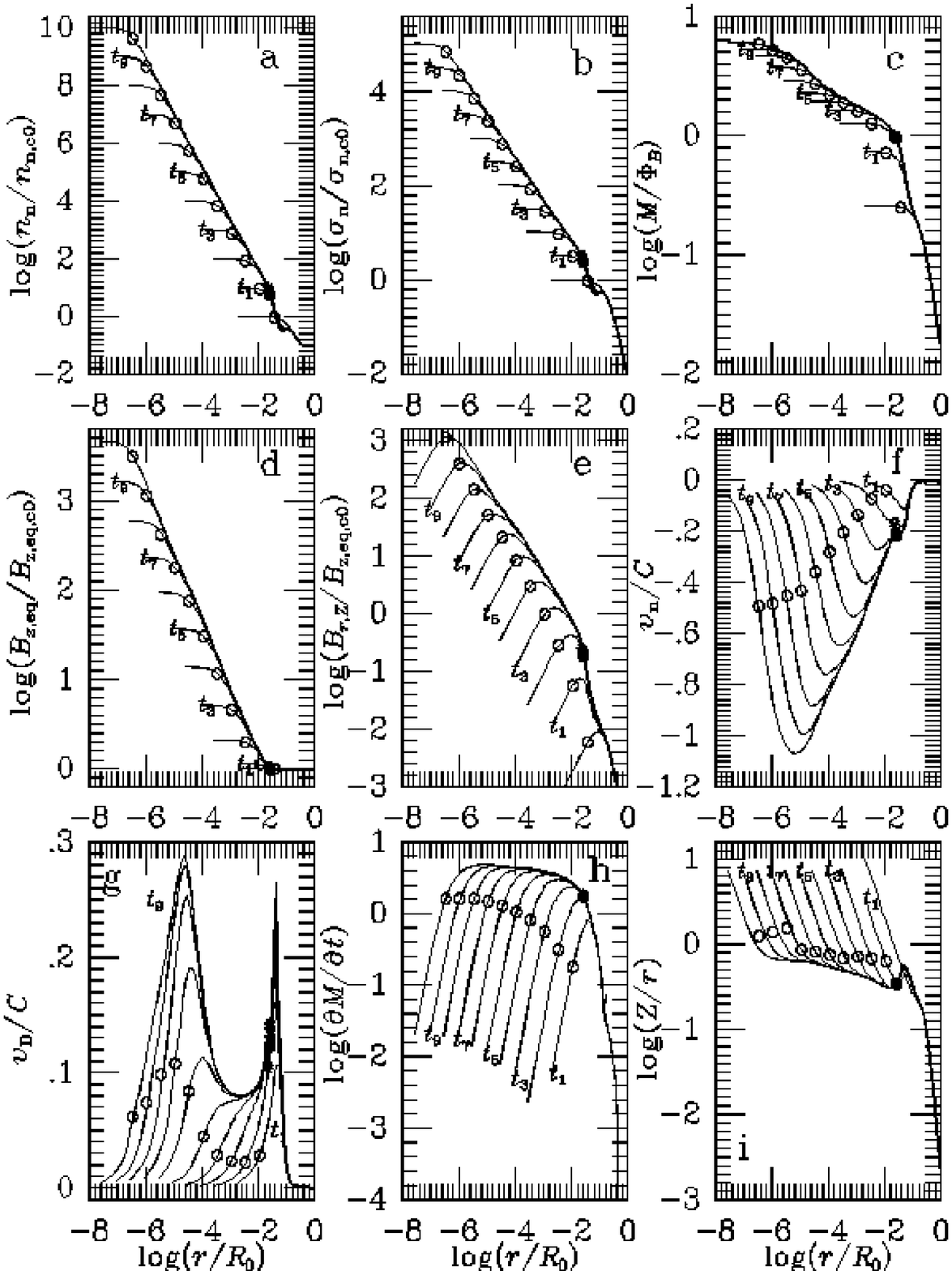}
\caption{}
\end{figure}
\begin{figure}
\plotone{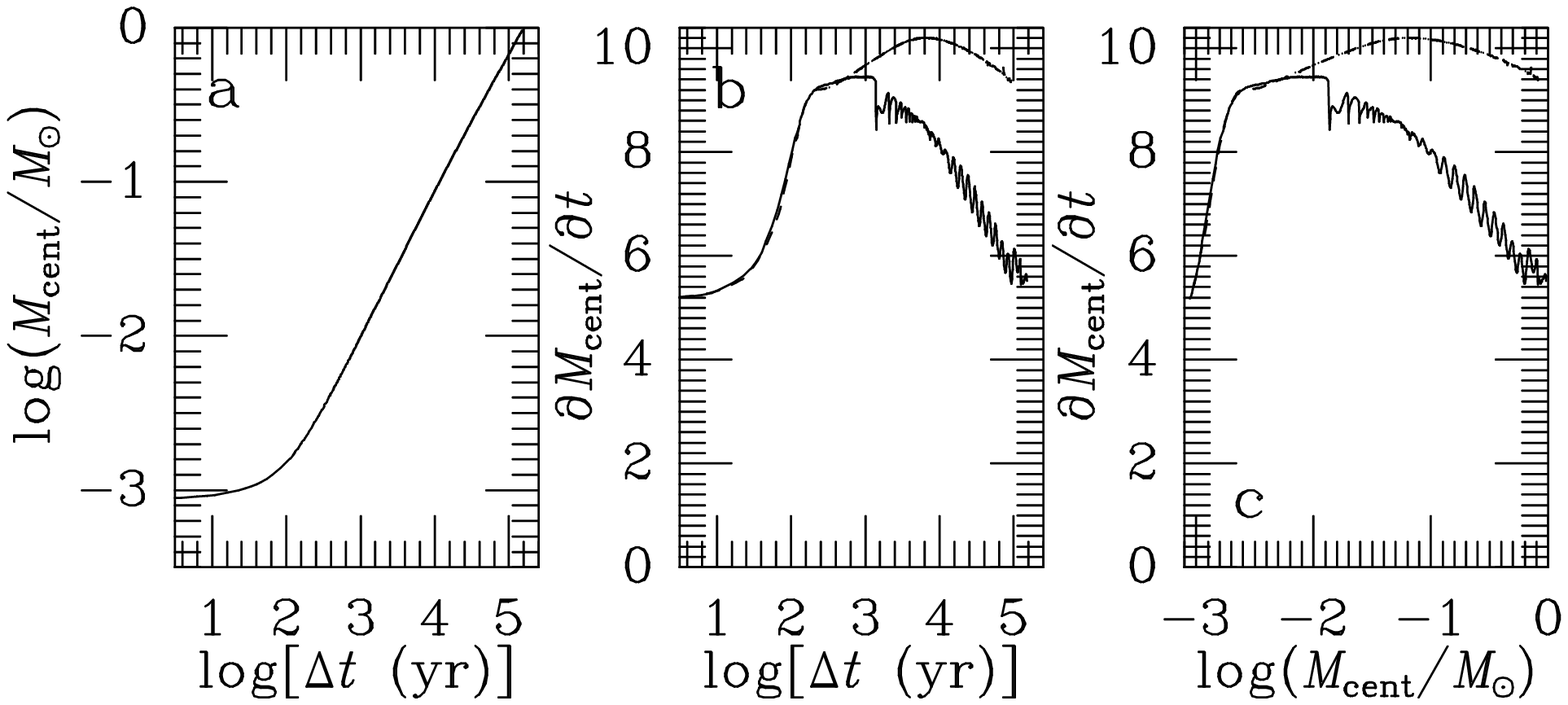}
\caption{}
\end{figure}
\begin{figure}
\plotone{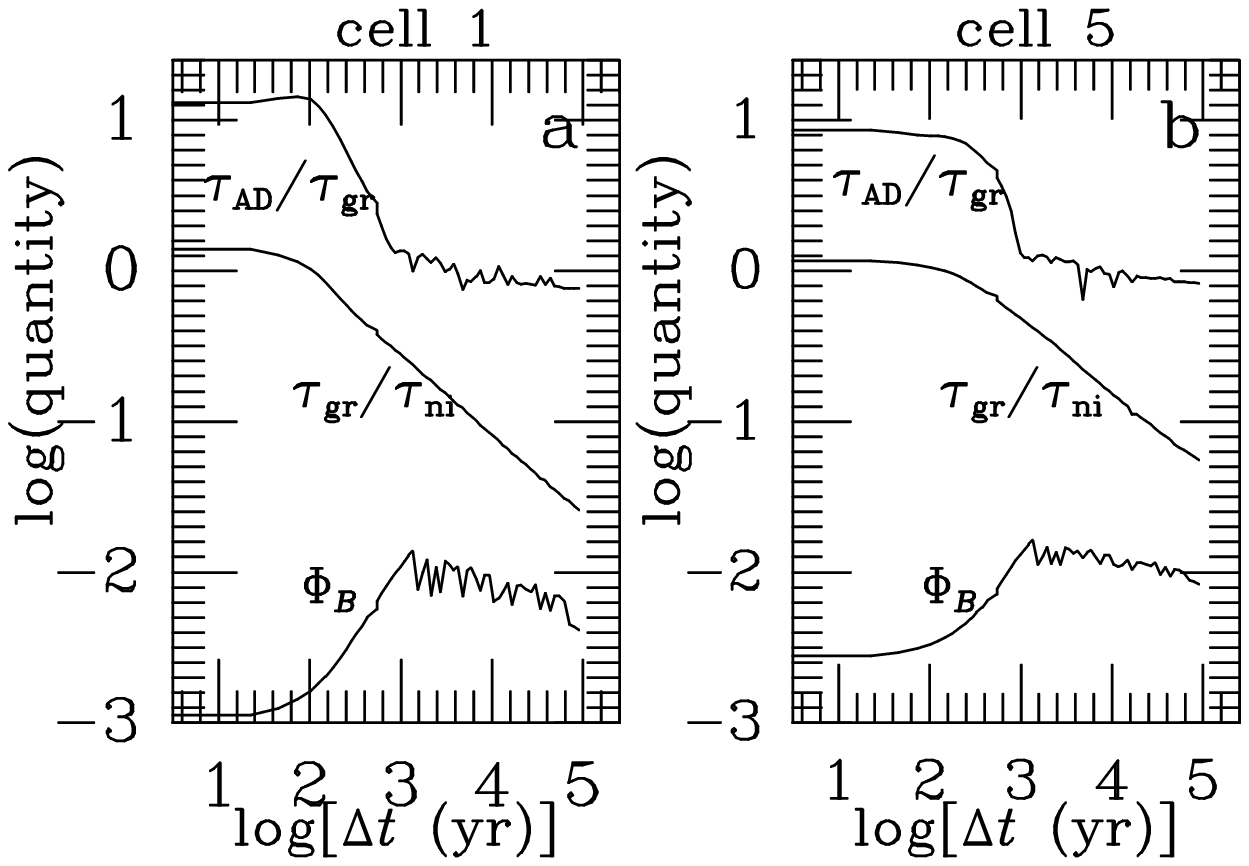}
\caption{}
\end{figure}
\begin{figure}
\plotone{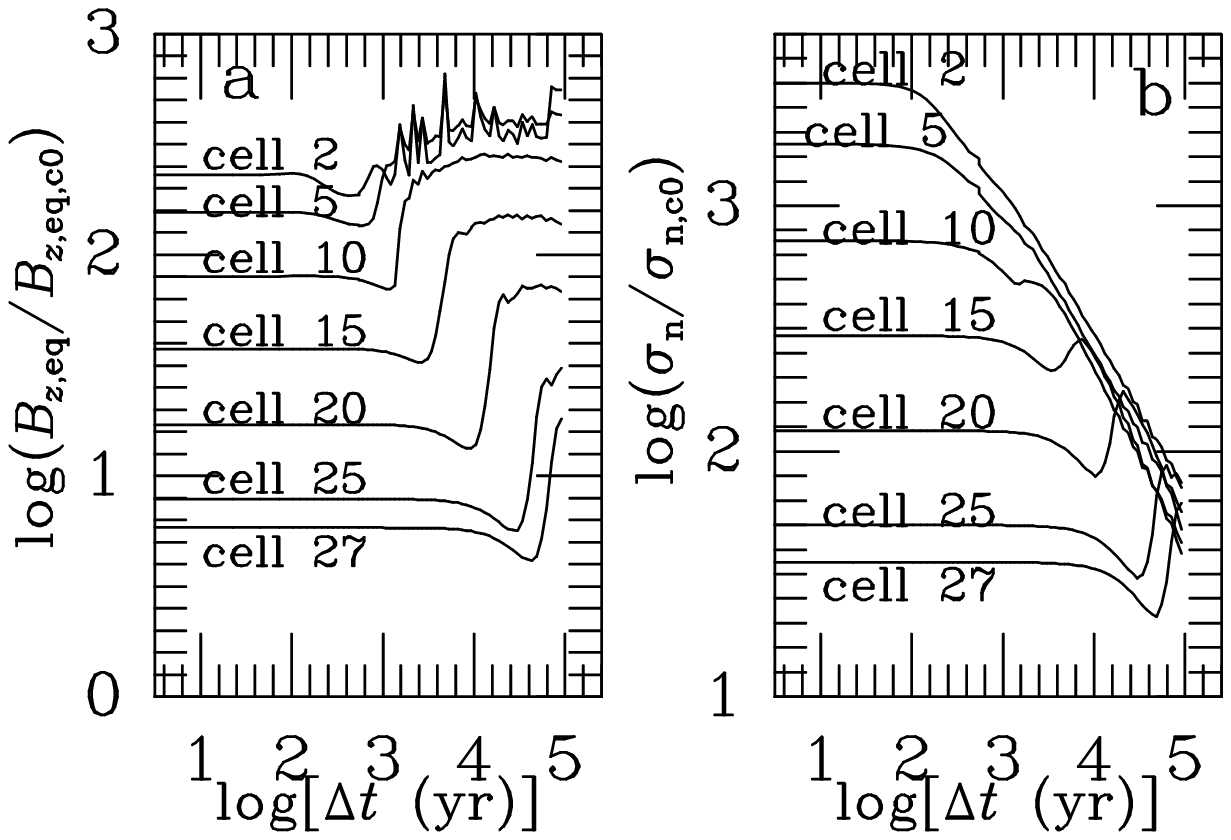}
\caption{}
\end{figure}
\begin{figure}
\plotone{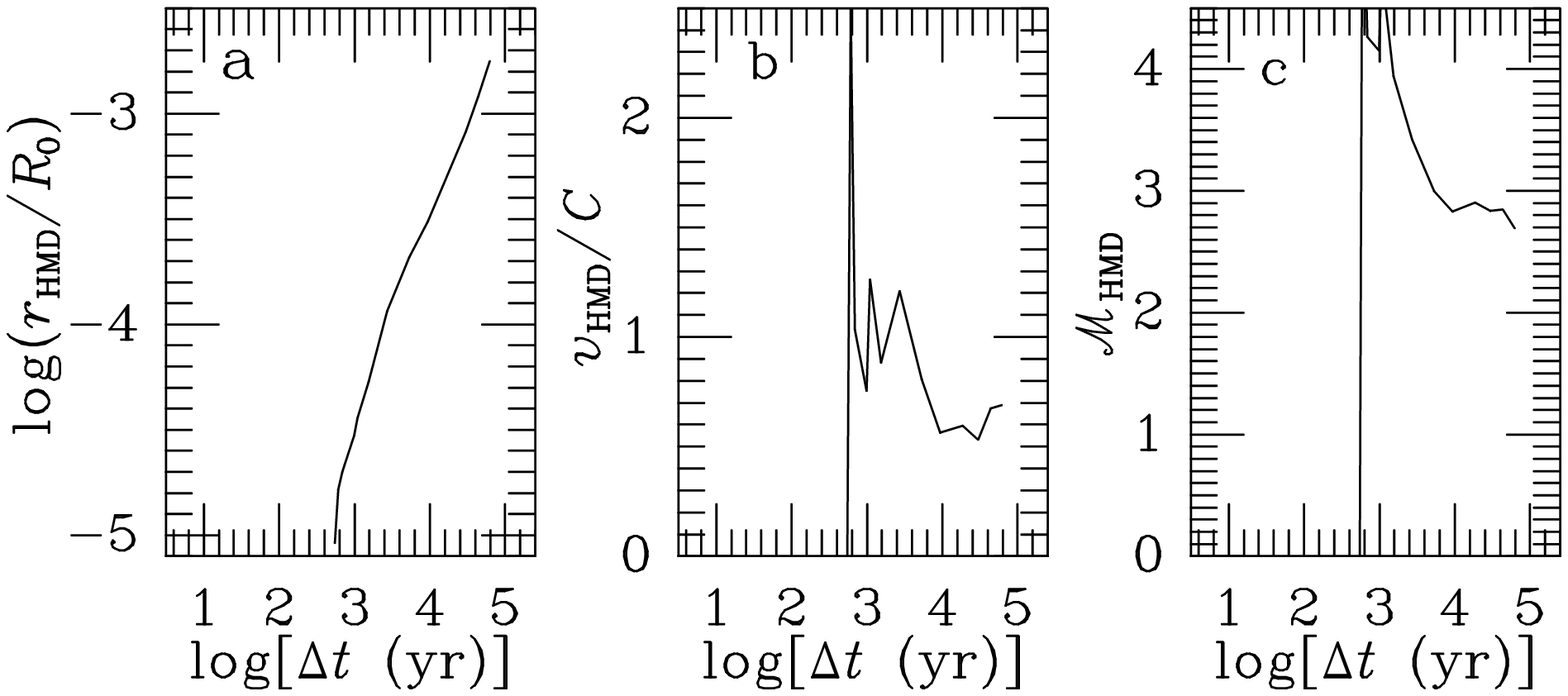}
\caption{}
\end{figure}
\begin{figure}
\plotone{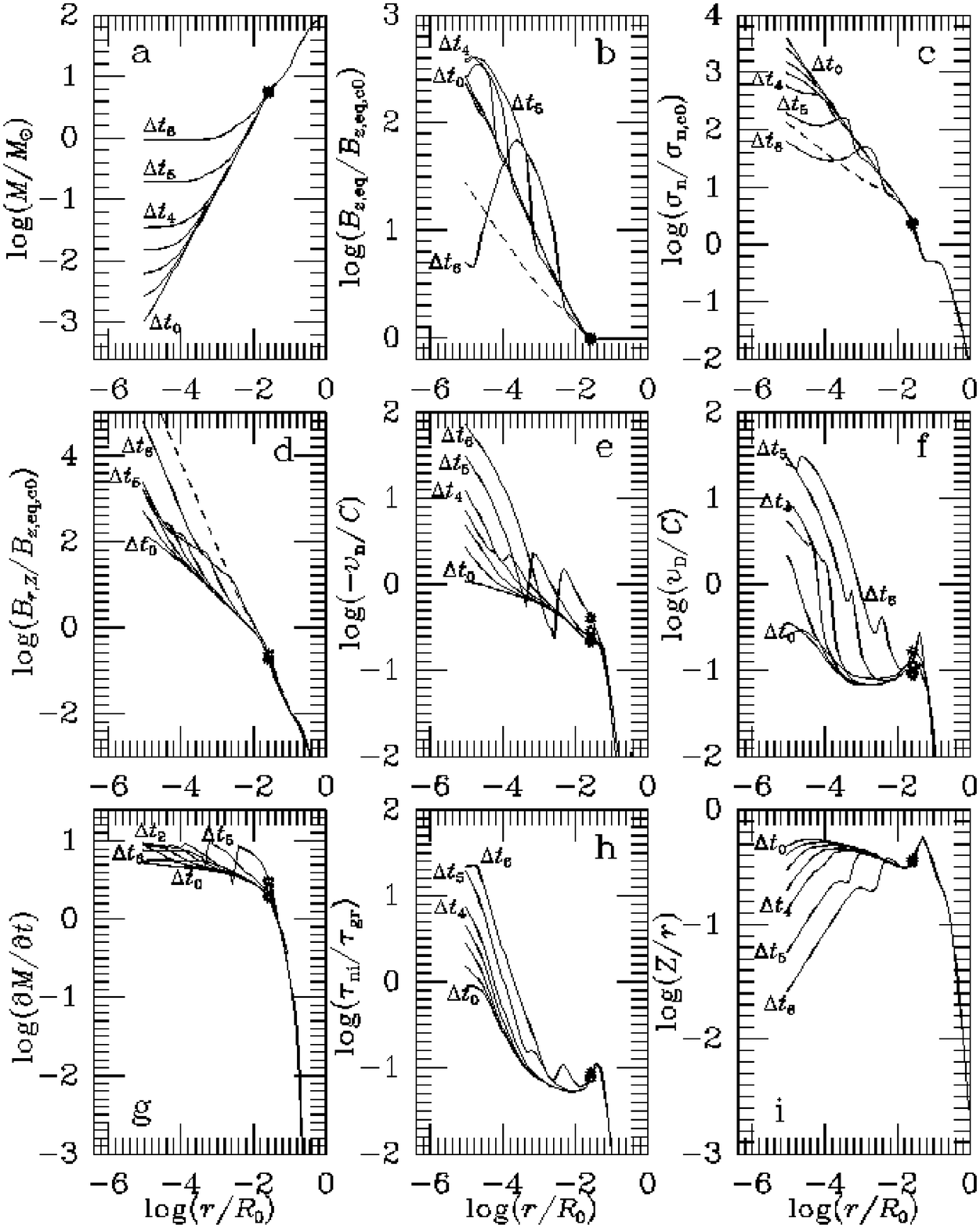}
\caption{}
\end{figure}
\begin{figure}
\plotone{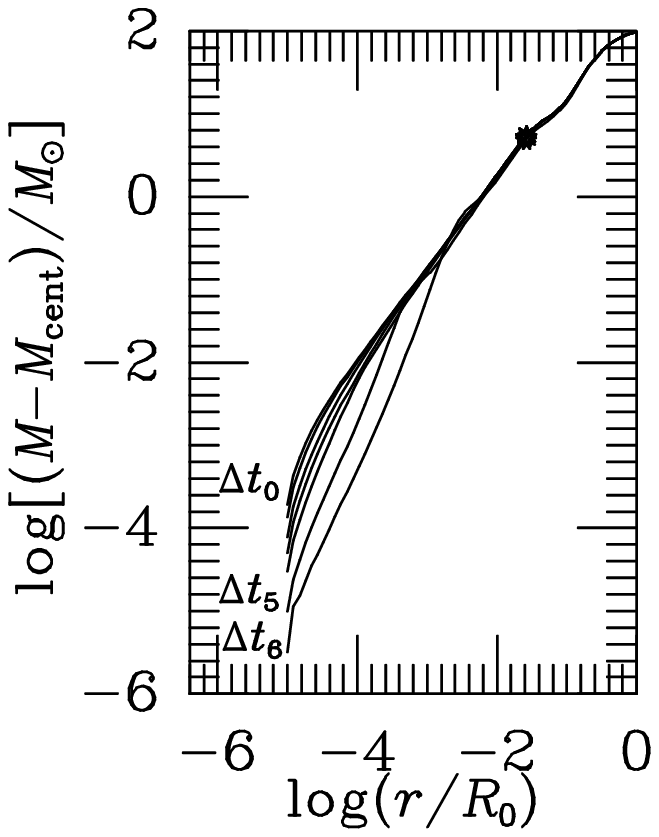}
\caption{}
\end{figure}
\begin{figure}
\plotone{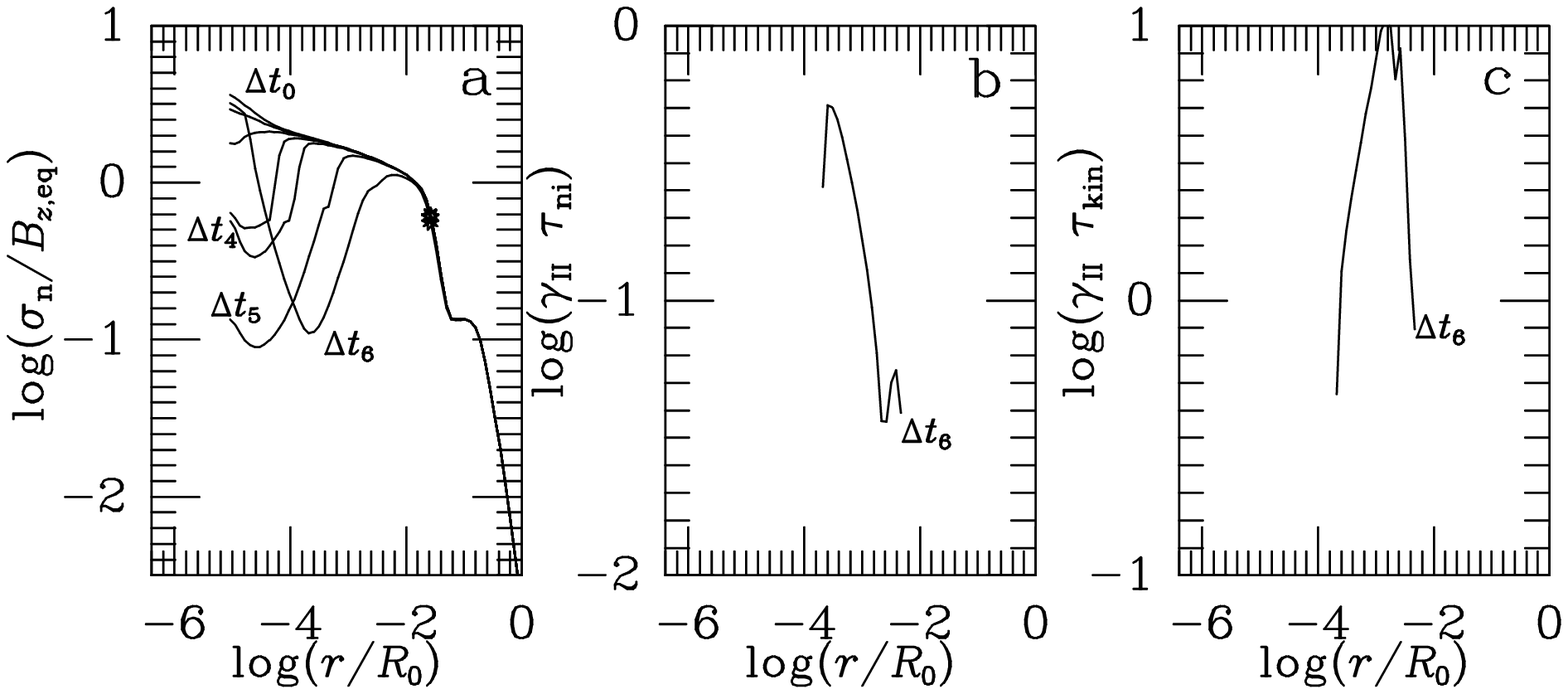}
\caption{}
\end{figure}
\begin{figure}
\plotone{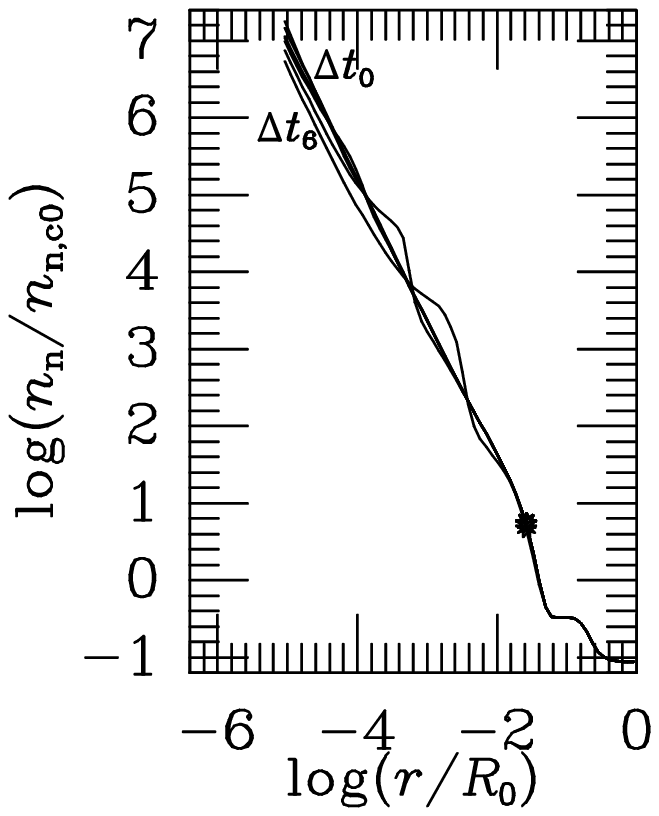}
\caption{}
\end{figure}
\begin{figure}
\plotone{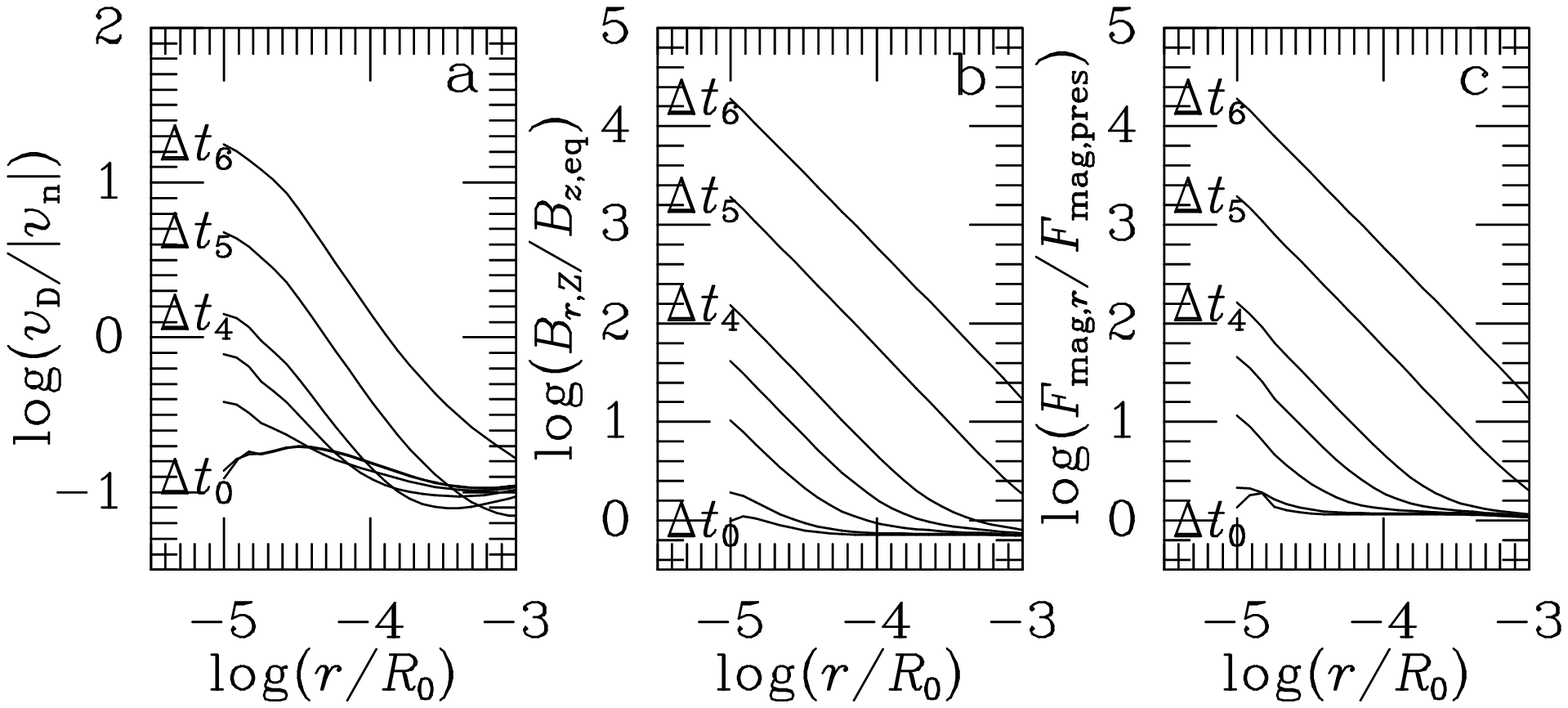}
\caption{}
\end{figure}

\end{document}